\documentclass[floatfix,showpacs,amsmath,amsfonts,amssymb,aps,twocolumn,superscriptaddress,pra,10pt]{revtex4-1}
\usepackage[pdftex]{graphicx}
\usepackage[pdftex]{hyperref}

\newcommand{\expe}{\mathrm{e}}
\newcommand{\du}{\mathrm{d}}
\newcommand{\br}{\mathbf{r}}
\newcommand{\bR}{\mathbf{R}}
\newcommand{\bk}{\mathbf{k}}
\newcommand{\bx}{\mathbf{x}}
\newcommand{\bq}{\mathbf{q}}
\newcommand{\bzero}{\mathbf{0}}
\newcommand{\ii}{\mathrm{i}}

\newcommand{\ket}[1]{| #1 \rangle}
\newcommand{\avg}[1]{\langle #1 \rangle}
\newcommand{\kF}[0]{k_\mathrm{F}}
\newcommand{\EF}[0]{E_\mathrm{F}}
\newcommand{\rc}[0]{r_\mathrm{c}}
\newcommand{\Reff}{R_\mathrm{eff}}
\newcommand{\bigo}{O}
\newcommand{\mytimes}{\times}
\DeclareMathOperator{\arccot}{arccot}

\newcommand{\figref}[1]{Fig.~\ref{#1}}
\newcommand{\appref}[1]{Appendix~\ref{#1}}
\newcommand{\reference}[1]{Ref.~\cite{#1}}
\newcommand{\refs}[1]{Refs.~\cite{#1}}
\newcommand{\secref}[1]{Section~\ref{#1}}
\newcommand{\eqnref}[1]{Equation~(\ref{#1})}

\begin{document}

\title{Effective-range dependence of resonant Fermi gases}
\author{L.M.~Schonenberg}
\affiliation{Cavendish~Laboratory, J.J.~Thomson~Avenue, Cambridge, CB3~0HE, United Kingdom}
\author{G.J.~Conduit}
\affiliation{Cavendish~Laboratory, J.J.~Thomson~Avenue, Cambridge, CB3~0HE, United Kingdom}
\date{\today}

\begin{abstract}
A Fermi gas of cold atoms allows precise control over the
dimensionless effective range, $\kF \Reff$, of the Feshbach
resonance. Our pseudopotential formalism allows us to create smooth
potentials with effective range, $-2 \leq \kF \Reff \leq 2$, which we
use for a variational and diffusion Monte Carlo study of the
ground state of a unitary Fermi gas. We report values for the
universal constants of $\xi = 0.388(1)$ and $\zeta = 0.087(1)$, and
compute the condensate fraction, momentum distribution, and pair
correlations functions. Finally, we show that a gas with $\kF \Reff
\gtrsim 1.9$ is thermodynamically unstable.
\end{abstract}

\maketitle

\section{Introduction}

Cold atom gases have delivered a series of surprises and insights,
including polaron physics \cite{Schirotzek2009}, the realization of
the Bose-Hubbard model \cite{Greiner2002}, and the BEC-BCS crossover
\cite{Leggett1980,Bourdel2004,Giorgini2008}. The development of
uniform trapping potentials has enabled the experimental realization
of particles in a box \cite{Gaunt2013}, while the Feshbach resonance
offers a unique level of control of the inter-particle interactions
\cite{Chin2010}. Fermi gases interacting via zero-range contact
interactions offer scale invariant physics in the unitary limit of
diverging scattering length, captured by the Bertsch parameter
\cite{Baker1999}. However, despite their universal physics, contact
interactions do not represent finite range interactions seen in
nature, e.g. screened Coulomb forces, neutron-neutron interactions,
and narrow Feshbach resonances. In this paper we present a study of
the consequences of finite ranged interactions in a unitary Fermi gas.

The scattering of two particles at low energies is described by the
scattering phase shift \cite{Landau1981}, which up to first order in
the wave vector $k$ is given by
\begin{equation*}
  \label{eq:phaseshift}
  \cot(\delta(k)) = -\frac{1}{k a} + \frac{1}{2} k \Reff,
\end{equation*}
where $a$ is the scattering length and $\Reff$ the effective range. In
the limit of zero interaction range $\Reff = 0$, a vanishing
scattering length $a=0$ corresponds to a noninteracting gas, while
the unitary limit of infinite scattering length $a^{-1}=0$ results
in scale-invariance. We use this scale-invariance as a solid basis to
investigate the effects of the length scale introduced by the
effective range term $\Reff$. Typical values for the effective range
are $\kF \Reff \approx 3$ \cite{Miller1990,Forbes2012} for
neutron matter, and $\kF \Reff \gtrsim -4$ for the $543.25 \mathrm{G}$
narrow Feshbach resonance of \textsuperscript{6}Li
\cite{Gurarie2007,Wang2011}. There is a wide variety of Feshbach
resonances available \cite{Chin2010} and several of those exhibit
large negative effective ranges, summarized in \reference{Wang2011}.

So far most quantum Monte Carlo (QMC) studies of finite-range interactions have used
the P\"oschl-Teller interaction potential for $0<\kF \Reff<0.4$, and then
extrapolate effective range effects to zero to study the ground state
of the unitary Fermi gas \cite{Li2011,Gandolfi2011,Forbes2011}. 
\citet{Forbes2012} purposefully consider the effect of small
positive effective ranges up to $\kF \Reff = 0.35$ in the context of
neutron matter. Negative effective ranges have been studied in an
Eagles-Leggett mean-field theory using a well-barrier interaction
potential at zero temperature \cite{DePalo2004,Jensen2006}, at
finite-temperature \cite{DePalo2005}, and also using the two-channel
model of the Feshbach resonant interaction at both zero and finite
temperatures \cite{Timmermans1999,Gurarie2007}.

Here we study a gas at unitarity across a broad spread of effective
ranges $-2 \leq \kF \Reff \leq 2$. Many-body physics arises from
repeated two-body scattering events, so a Hamiltonian where the
opposite spin fermions interact via a pseudopotential that exactly
reproduces the scattering phase shift with $a^{-1}=0$ and $\Reff$ is
the ideal platform for an accurate many-body simulation. To smoothly
connect positive and negative effective ranges we develop a new
pseudopotential following \refs{Bugnion2014,Whitehead2016a}. The
pseudopotential is smooth and extended in space, making it easy to
sample with the variational and diffusion Monte Carlo methods that we
use to calculate ground state properties \cite{Foulkes2001}.

In \secref{formulation} we use two-body scattering theory to
understand the properties of our potential. In
\secref{pseudopotentials} we evaluate four possible choices for the
interaction potential, including the newly proposed Ultra-Transferable
Pseudopotential (UTP) and select the UTP as the potential of choice
for numerical studies. In \secref{qmc} we discuss the quantum Monte
Carlo formalism and present results for the ground state energy
including values for the universal constants, condensate fraction,
momentum distribution and Tan's contact, and pair correlation
functions. Finally, we consider the thermodynamic stability of the
system in \secref{stability} and find that gases with $\kF \Reff
\gtrsim 1.9$ are unstable.

\section{Formulation of the problem}
\label{formulation}

We study the Hamiltonian for spin $1/2$ fermions in three dimensions with resonant
interactions between opposite spins,
\begin{equation*}
 \label{eq:hamiltionian} 
\hat{H} = -\frac{1}{2} \sum_{i=1}^N \nabla_i^2 +
\sum_{i \neq j}^{N} V(r_{ij}).
\end{equation*} 
Atomic units ($\hbar = m = 1$) are used throughout. $\nabla_i^2$ is
the Laplacian with respect to the coordinates of particle $i$, $N$ is
the total number of particles and we study equal numbers of up and down
spin particles. $r_{ij}$ is the distance between particles $i$ and $j$, and $V$
is an interaction potential that acts between particles with opposite
spins, characterized by the idealized scattering phase shift
$\cot(\delta(k)) = k \Reff/2$.  To understand the form of this
interaction potential we first summarize some important results from
scattering theory and in particular consider the possible emergence of
bound states, before we discuss the explicit forms of the potentials
used.

\subsection{Scattering theory}
We consider two identical distinguishable fermions in a vacuum.  In
their center-of-mass frame, the Schr\"odinger equation for particles
interacting via a radially symmetric potential $V(r)$ is
given in spherical coordinates by
\begin{equation*}
\label{eq:Schrodinger}
[-\nabla^2 + V(r)] \psi(r,\theta,\phi) = E \psi(r,\theta,\phi) ,
\end{equation*}
where $E$ is the energy of the relative motion.

The analytic solution for noninteracting
particles, $V(r)=0$, takes the form
\begin{equation*}
\psi_{lm}(r,\theta,\phi) = Y_{lm}(\theta,\phi) R_l (r),
\end{equation*}
with $l$ the angular momentum and $m$ the component of the angular
momentum along the quantization axes. $Y_{lm}$ are the spherical
harmonics, and the radial function $R_l$ is given by
\begin{equation}
  \label{eq:radial}
  R_l(r) = \mathcal{A}_l(k) j_l(kr) + \mathcal{B}_l(k) n_l(kr),
\end{equation}
where $k=\sqrt{E}$ is the wave vector in the center-of-mass frame, and
the coefficients $\mathcal{A}_l(k)$ and $\mathcal{B}_l(k)$ are set by the
boundary conditions. $j_l(kr)$ and $n_l(kr)$ are the spherical Bessel
and spherical Neumann functions respectively. To connect to scattering
waves we rewrite the radial function in terms of spherical
Hankel functions $h_l^{(1,2)} (kr) = j_l(kr) \pm \ii n_l(kr)$,
\begin{equation*}
  R_l(r) = \mathcal{A}_l'(k) h_l^{(1)} (kr) + \mathcal{B}_l'(k) h_l^{(2)} (kr).
\end{equation*}
The Hankel functions $h_l^{(1,2)} (kr)$ behave as spherical waves at
large radii $\sim \exp[\pm \ii (kr-l\pi/2)]/r$.

The effect of a spherically symmetric interaction potential $V(r)$ on
the wave function is limited by angular momentum conservation
and causality to the introduction of a phase shift $\delta_l(k)$
in the outgoing wave $h_l^{(1)} (kr)$ of the radial wave function,
\begin{equation}
  \label{eq:scattering-wave function}
   R_l^\mathrm{int}(r) = \mathcal{C}_l(k) [\expe^{\ii 2 \delta_l(k)} h_l^{(1)} (kr) + h_l^{(2)} (kr)],
\end{equation} 
with $\mathcal{C}_l(k)$ a normalization constant. At large radii
$R_{l}^\mathrm{int} (r) \sim \sin(kr+\delta_l(k)-l\pi/2)/r$, verifying
the interpretation of $\delta_l(k)$ as a phase shift. $\delta_l(k)$ is
related to the coefficients in \eqnref{eq:radial} as
$\delta_l(k)=\arctan[-\mathcal{B}_l(k)/\mathcal{A}_l(k)]$. The ratio
$-\mathcal{B}_l(k)/\mathcal{A}_l(k)$ can be expressed in terms of the
logarithmic derivative of the interacting radial wave function by
matching $R_l^\mathrm{int}(r)$ and $R_l(r)$ at the cutoff radius $\rc$
beyond which the interaction potential vanishes. Combining both
results, the phase shift can be expressed as
\begin{equation*}
  \delta_l(k)=\arctan \bigg[\frac{k j_l'(k \rc)-\gamma_l j_l(k \rc)}
  {k n_l'(k \rc)-\gamma_l n_l(k \rc)} \bigg],
\end{equation*}
where $\gamma_l=(R_l^{\mathrm{int}})'(\rc)/R_l^\mathrm{int}(\rc)$.

At large radii, the interacting wave function can also be written as
the sum of an incoming plane wave and a spherical outgoing scattered
wave
\begin{equation*}
  \label{eq:scattering-wave function2}
  \lim_{\br \to \infty}\psi^\mathrm{int}(\br) = \expe^{\ii \bk \cdot \br} +
  \frac{f(k,\theta)}{r}\expe^{\ii kr},
\end{equation*}
with $f$ the scattering amplitude and $\theta$ the scattering
angle. By equating the radial component of this expression in angular
momentum channel $l$ with \eqnref{eq:scattering-wave function}, the
scattering amplitude can be related to the phase shift
\begin{equation*}
  \label{eq:scatteringamplitude-generic}
  f_l(k) = \frac{1}{\cot(\delta_l(k)) - \ii k}.
\end{equation*}
This expression reveals bound states of the interaction potential
because they introduce poles into the scattering amplitude $f_l(k)$
\cite{Landau1981}. From now on we focus on the $l=0$ channel that
dominates interactions between opposite spin fermions, starting with
an examination of possible bound states in the next section.

\subsection{Bound states}
\label{bound-states}

\begin{figure}
\includegraphics[width=\linewidth]{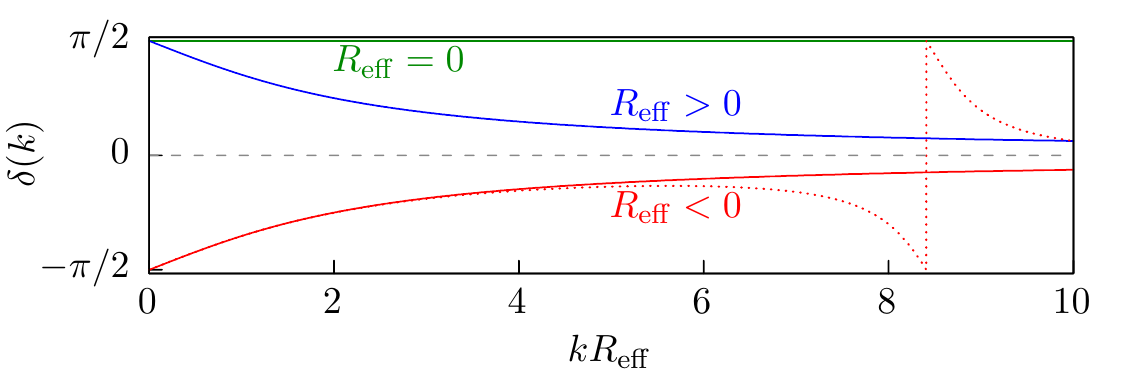}
\caption{(Color online) Scattering phase shift $\delta(k) =
\arccot(k \Reff/2)$ for the three different cases of $\Reff$. For
$\Reff < 0$ the phase shift of a realistic potential with the same low
energy scattering properties is indicated by the red dotted line. The
noninteracting phase shift is shown by the gray dashed line.}
\label{fig:scattering}
\end{figure}

Each time the phase shift accumulates a factor of $\pi$, a node is
introduced in the wave function of the scattered wave $\sim \sin(k r +
\delta(k))/r$. Since each node introduced into the wave function by the
potential corresponds to an additional bound state, this establishes
the link between the scattering phase shift and the number of bound
states $n \geq 0$ for any well-behaved potential, which is formalized
in Levinson's theorem \cite{Chadan1989},
\begin{align*}
\delta(0) - \delta(\infty)=\begin{cases}
n \pi,& a^{-1} \neq 0, \\
(n+\frac{1}{2}) \pi,& a^{-1} = 0.
\end{cases}
\end{align*}

We are interested in the latter case $a^{-1} = 0$.  As is evident from
\figref{fig:scattering}, for $\Reff > 0$ the phase shift decreases
from $\pi/2$ to $0$, and there is no bound state. For $\Reff < 0$,
$\delta(0) - \delta(\infty) = -\pi/2$, which gives $n = -1$. Because
the number of bound states cannot be negative, this phase shift does
not correspond to a physical potential. However, potentials with the
same low-energy scattering properties may be obtained from a phase
shift with additional contributions at higher order in $k$. Provided
these contributions occur at momenta beyond the largest momentum scale
in the system, i.e. the Fermi momentum $\kF$ for a fermionic many-body
system, they do not affect the physics of the system as the
interacting particles cannot probe these high momentum features. As
seen in the figure, the effect of the higher order term is to
introduce a phase winding of $\pi$ so that $\delta(0) - \delta(\infty)
= \pi/2$, which corresponds to a physical potential with no bound
state. We conclude that in both cases there is no bound state and
the potential is therefore completely characterized in terms of its
scattering phase shift \cite{Chadan1989}.

Despite the absence of a true bound state with negative energy,
virtual bound states may exist. The scattering amplitude for our
idealized phase shift reads
\begin{equation*}
  \label{eq:bindingenergy}
  f = \frac{1}{\frac{1}{2} k^2 \Reff - \ii k},
\end{equation*}
which has a pole at zero energy $k = 0$. In the zero-range limit,
$\Reff=0$, this pole corresponds to a zero energy virtual bound state,
which is the $a \to \infty$ limit of the familiar bound state with
energy $E=-1/(2a^2)$ \cite{Giorgini2008,Gurarie2007}. Because the pole
in the scattering amplitude extends to finite effective range
$\Reff$, so does the virtual bound state, which will be important for
our discussion of the many-body system in \secref{energy}.

\section{Pseudopotentials}
\label{pseudopotentials}

Having defined the interaction potential in terms of scattering
properties, we evaluate four possible real space interaction
potentials for use in our many-body simulations. For positive
effective range we consider the potential well and P\"oschl-Teller
interactions; for negative effective range we consider the
well-barrier potential. Furthermore, we propose the Ultra-Transferable
Pseudopotential (UTP)
\cite{Bugnion2014,Lloyd-Williams2015,Whitehead2016,Whitehead2016a},
which is equally applicable for both positive and negative effective
ranges. After comparing all four potentials, we select the UTP for our
numerical study. The software used to generate the UTP is
available online \cite{Schonenberg2016a}.

\subsection{Positive effective range}

Positive effective ranges $\Reff>0$ for attractive interactions
are usually obtained from uniformly attractive potentials, $V(r) \leq 0$
for all $r$. In this case, the effective range
is approximately equal to the physical interaction range \cite{Landau1981}, while
the depth of the potential can be used to tune the scattering length.

\subsubsection{Potential well}

A spherical potential well interaction was used in
\refs{Astrakharchik2004,Astrakharchik2005,Jensen2006} as a model for
contact interactions,
\begin{align*}
V(r)=\begin{cases}
-\bigg(\dfrac{\pi}{2\Reff}\bigg)^2,& r\le \Reff, \\
0,& r>\Reff,
\end{cases}
\end{align*}
tuned to have scattering length
$a^{-1}=0$, and effective range $\Reff$. The scattering phase shift of
this potential is correct at low incident energies, but is incorrect
at intermediate energies where higher order terms start to contribute.

\subsubsection{P\"oschl-Teller}

The P\"oschl-Teller interaction gives the exact phase shift with
scattering length $a$ and effective range $\Reff$ in the lowest
angular momentum channel \cite{Chadan1989}, and has been
used in several studies
\cite{Morris2010,Carlson2011,Forbes2012,Li2011}. At unitarity,
$a^{-1}=0$, the potential can be written in terms of its effective
range $\Reff$ as
\begin{equation*}
  V(r) = -\frac{8 \Reff^{-2}}{\cosh^2 (\frac{2 r}{\Reff})}.
\end{equation*}

\subsection{Negative effective range}

Scattering phase shifts with negative effective range result from
potentials with an attractive well hosting a (virtual) bound state at
short radii, and a potential barrier at intermediate radii. Quantum
tunneling through the potential barrier couples the (virtual) bound
state with the continuum of scattering states at large radii. When a
rising barrier suppresses quantum tunneling, the (virtual) bound and
scattering states become uncoupled.

These potentials are called Shape resonances and exhibit the same
physics as Feshbach resonances. In the Feshbach resonance model the
(virtual) bound state in the well is represented by the closed
channel, and the tunneling through the potential barrier is described
by a hybridization term that mixes the closed channel with the open
channel that describes the continuum of scattering states \cite{Wang2011}.

\subsubsection{Well-barrier potential}

Following \refs{DePalo2004,DePalo2005,Jensen2006} we consider 
a well-barrier potential,
\begin{align*}
V(r)=\begin{cases}
-U_0,& r\le R_0, \\
U_1,& R_0 < r \le R_1, \\
0,& r>R_1,
\end{cases}
\end{align*}
with $U_0, U_1 > 0$ and $R_1 > R_0 > 0$. This potential reduces to the
potential well for $U_1=0$. A potential with scattering length $a$ and
effective range $\Reff$ for given radii $\{R_0,R_1\}$ can be obtained by suitably tuning the well
depth and barrier height $\{U_0,U_1\}$ as described in
\reference{Jensen2006}. We discuss our choice for $\{R_0,R_1\}$ in \secref{comparison}.

As discussed in \secref{bound-states}, the scattering phase shift of
physical potentials with negative effective range include a phase
winding by $\pi$ at some high momentum $k$. Dimensional analysis confirms
that this momentum may be pushed to arbitrarily high momentum where it
does not affect the scattering of low-energy particles by reducing
$\{R_0, R_1\}$, at the expense of diverging $\{U_0, U_1\}$.

\subsection{UTP}
\label{UTP}

We now propose a pseudopotential that describes both positive and
negative effective ranges. It is also smooth and extended in space,
easing the application of numerical methods. Following
\cite{Bugnion2014,Whitehead2016a} we propose a UTP that takes a
polynomial form within a cutoff radius $\rc$,
\begin{align*}
V^\mathrm{UTP\!}(r)\!=\!\begin{cases}
\!\left(\!1\!-\!\frac{r}{\rc}\!\right)^2 \!\left[ u_1\!\left( 1+\frac{2r}{\rc} 
\right)\! +\! \displaystyle\sum_{i=2}^{N_\mathrm{u}} u_i 
\left(\!\frac{r}{\rc}\!\right)^{\!i} \right]\!,&\!\!\!\!r\le\rc,\\
0,&\!\!\!\!r>\rc,
\end{cases}
\end{align*}
where the $u_i$ are the $N_\mathrm{u}=5$ optimizable coefficients. The
term $(1-r/\rc)^2$ ensures that the pseudopotential goes smoothly to
zero at $r=\rc$, and the component $u_1(1+2r/\rc)$ constrains the
pseudopotential to have zero gradient at particle coalescence,
to ensure that the wave function in the potential is smooth.

We calibrate the potential to deliver the correct scattering phase
shift for particles with momenta up to the characteristic momentum
scale of our many-body system, the Fermi momentum $\kF$. To determine
the coefficients $\{u_i\}$ we numerically solve the scattering
problem, extract the scattering phase shift
$\delta_l^\mathrm{UTP}(k)$, and then minimize the total squared error
in the phase shift over angular momentum channels $l$ and relative
scattering wave vectors $k$ of particles in the Fermi sea,
\begin{align*}
\label{eq:rmsphaseshift}
\langle \left| \delta_l^\mathrm{UTP}(k) - \delta_l(k) \right|^2 
\rangle&= \\ \nonumber
\sum_l \int_0^{\kF}&\left| \delta_l^\mathrm{UTP}(k) - \delta_l(k) \right|^2 
g(k/\kF) \du k,
\end{align*}
where the weighting is given by the density of scattering wave vectors
in the center of mass frame \mbox{$g(x)=12 x^2(2-3x+x^3)$}. The virtue
of a large cutoff radius $r_\mathrm{c}$ is that it leads to
potentials that are more extended in space. On the other hand,
$r_\mathrm{c}$ should be smaller than the inter-particle spacing so
that three-body scattering events are rare.  We therefore choose $r_c
= 1/\kF$, except for large positive effective ranges where we need a
cutoff radius of the order of $\Reff$, so we adopt $r_c = \max(1/\kF ,
2\Reff)$.

\subsection{Comparison of potentials}
\label{comparison}

\begin{figure}
\includegraphics[width=\linewidth]{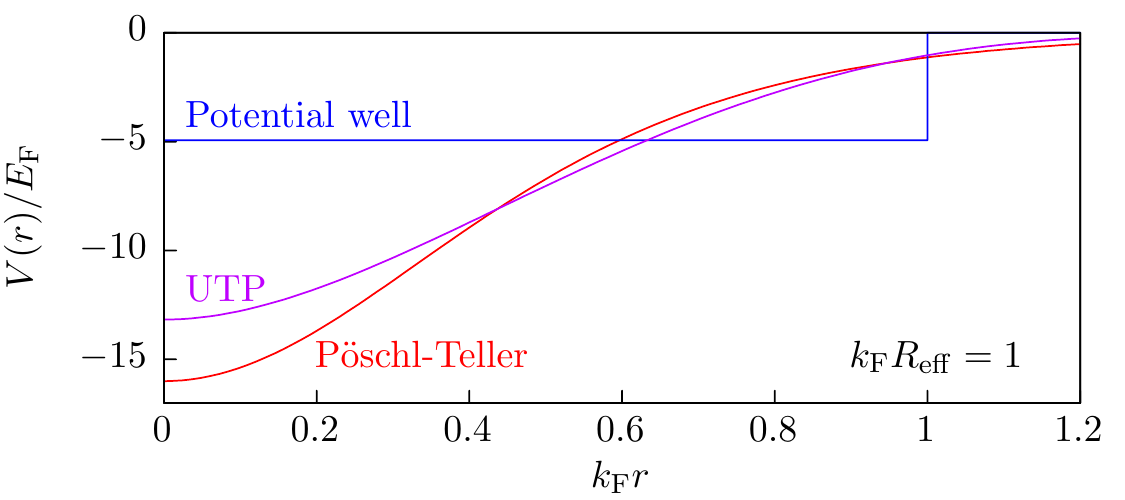}
\includegraphics[width=\linewidth]{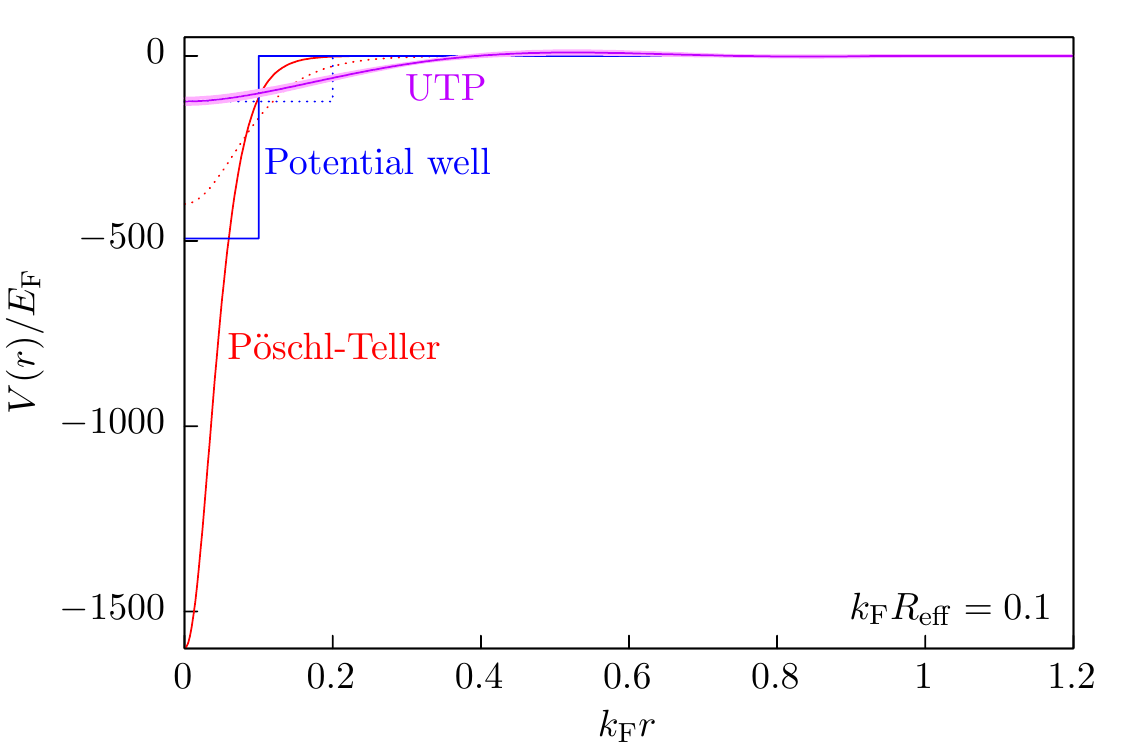}
\includegraphics[width=\linewidth]{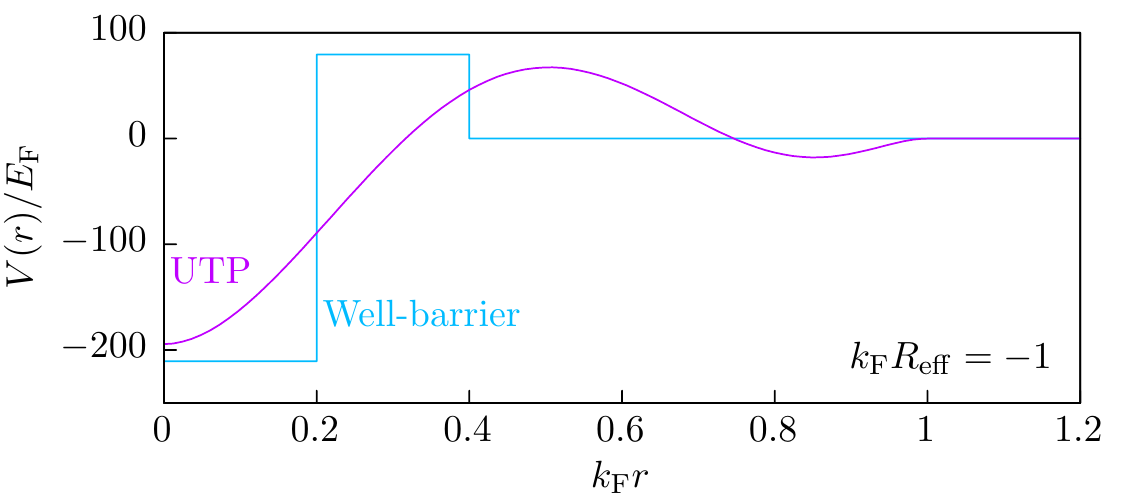}
\caption{(Color online) Plots of different potentials normalized by the
  reciprocal Fermi energy $\EF$ as a function of radius for
effective ranges $\kF \Reff = \{-1,0.1,1\}$. For the case $\kF \Reff =
0.1$, the potentials are shown using solid lines, whereas the red and
blue dashed lines show the P\"oschl-Teller and potential well
interactions with $\kF \Reff = 0.2$, and the shaded purple region
shows the variation of the UTP for effective range $0 \leq \kF \Reff
\leq 0.2$. The parameters of the well-barrier potential potential were
chosen so that its depth and height correspond to those of the UTP.}
\label{fig:potentials}
\end{figure}

Having introduced four possible interaction potentials, we now compare
how accurately they recover the correct scattering phase shift and their
numerical efficiency. The latter is a combination of two
factors: numerical convergence is aided by smooth potentials as
they produce smooth wave functions, and also by potentials with a wide
spatial extent as they occupy a larger volume of configuration space
so are more rapidly sampled.

To visualize the smoothness and extent of the interaction potentials
we plot potentials with effective ranges $\kF \Reff = \{-1,0.1,1\}$ in
\figref{fig:potentials}. For positive effective range the UTP is
similar to the P\"oschl-Teller interaction. The potential well is of
similar spatial extent, but shows a discontinuity. The diverging depth
of the P\"oschl-Teller and potential well interactions in the
zero-range limit, $\kF \Reff = 0$, is illustrated by the deepening of
those potentials as the effective range decreases from $\kF
\Reff=0.2$, indicated by the dotted line, to $\kF \Reff=0.1$,
indicated by the solid line. In this limit the interaction becomes
momentum independent, which for the potential well and P\"oschl-Teller
interactions implies that they also become short-ranged. This is not
the case for the UTP, as we calibrate the potential only for wave
vectors up to an intermediate momentum scale $\kF$. The shaded region
shows the variation of the UTP with effective range $0 \leq \kF \Reff
\leq 0.2$, demonstrating its shape remains similar even in the
zero-range limit. The numerical advantage of the UTP becomes clear: it
remains smooth and extended in space. For negative effective range the
UTP displays a barrier at intermediate distances, like the
well-barrier potential. For the well-barrier potential we set $\{\kF
R_0=0.2,\kF R_1=0.4\}$, so that its depth and height are similar to
that of the UTP. Many-body simulations will benefit from the
smoothness of the UTP compared to the two discontinuities for the
well-barrier. We conclude that the UTP is the only potential that is
of finite depth at all effective ranges, is smooth and extended in space,
and is therefore well-suited for use in a QMC simulation.

\begin{figure}
\includegraphics[width=\linewidth]{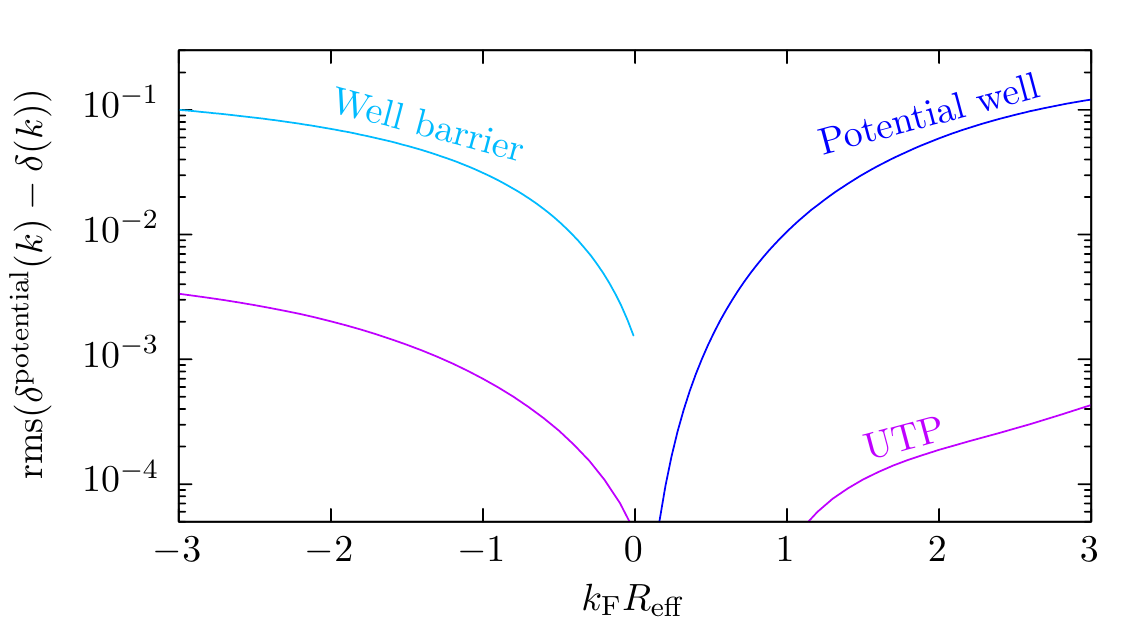}
\caption{(Color online) Root Mean Square (RMS) phase shift error for
the potential well, well-barrier and UTP for $k$ ranging from $0$ to
$\kF$. On average, the RMS
phase shift error for the UTP is about two orders of magnitude less
than that for the potential well and well-barrier potentials.}
\label{fig:phase_shift}
\end{figure}

Having examined the numerical advantages of the UTP compared to the
other potentials, we now evaluate the accuracy of their
scattering phase shifts. In \figref{fig:phase_shift} we plot the root
mean square (RMS) phase shift error of the potentials summed over
angular momentum channels and integrated over
scattering wave vectors up to the Fermi wave vector $\kF$. For
positive effective range the UTP is over two orders of magnitude more
accurate than the potential well. Its maximum error of less than
$10^{-3}$ also renders it equivalent to the exact P\"oschl-Teller
interaction for all practical purposes. For negative
effective ranges we note that even though the depth and height of the
well-barrier potential have been chosen to mimic the UTP, its phase
shift properties are almost two orders of magnitude worse. As the UTP
is the easiest to work with in a many-body simulation,
accurate, and applies at all effective ranges, we select it for our
QMC many-body study.

\section{Quantum Monte Carlo}
\label{qmc}

To calculate the ground state properties of the Fermi gas we use a
quantum Monte Carlo (QMC) method that is a tandem of the variational
Monte Carlo (VMC) and fixed-node diffusion Monte Carlo (DMC)
techniques \cite{Ceperley1980,Umrigar1993,Foulkes2001}. We use the
\textsc{casino} implementation \cite{Needs2010} with a Slater--Jastrow
trial wave function $\Psi = \expe^J D$, where $D$ is a Slater
determinant of $N/2$ pairing orbitals $\phi(\br_{ij})$, each holding
an up and down spin particle, and $J$ a Jastrow factor that we
optimize first using VMC, before using DMC to further relax the wave
function to its ground state. DMC is an accurate Green's function
projector method for determining ground state energies and other
expectation values, and is well-suited to investigating homogeneous
gaseous phases.

The pairing orbitals \cite{Carlson2003,Morris2010} are formed of a
linear combination of plane-waves and polynomials
\begin{align*}
  \phi(\br_{ij}) =& \sum_{n=0}^{N_\mathrm{PW}} a_n \sum_{\mathbf{G} \in
    S_n} \exp(\ii \mathbf{G} \cdot \br_{ij}) \\ \nonumber
  &+ \Theta(L_\mathrm{P} - r_{ij})
  \bigg(1-\frac{r_{ij}}{L_\mathrm{P}} \bigg)^3
  \sum_{n=0}^{N_\mathrm{P}} b_n r_{ij}^n,
\end{align*}
with $\mathbf{G}$ an element
of the set of symmetry related reciprocal lattice vectors $S_n$,
$N_\mathrm{PW}$ is the number of sets to include, and
$N_\mathrm{P}$ is the order of the polynomial. $\br_{ij}=\br_i-\br_j$
is the separation between two particles with
opposite spins at positions $\br_i$ and $\br_j$, and $r_{ij} =
|\br_{ij}|$ its magnitude. The term
$\Theta(L_\mathrm{P} - r_{ij}) (1-r_{ij}/L_\mathrm{P})^3$, where
$\Theta$ is the Heaviside step function and $L_\mathrm{P}$ is an
optimizable cutoff length, ensures that the polynomial orbital
smoothly approaches zero before the edge of the cell. The coefficients
$a_n$ and $b_n$ are optimizable, with the exception of $a_0$, which we
set to 1, and $b_1$, which is fixed by requiring that the orbital is
cuspless at the origin. The Slater determinant of these orbitals
contains both the noninteracting limit where the particles fill the
$N_\mathrm{PW}$ shells of plane-waves, and a superconducting state of
Cooper pairs captured by the polynomial series.

As superconductivity is a collective phenomenon, it is important to to
capture many-body correlations in the pairing orbitals. We therefore
use a backflow transformation \cite{LopezRios2006} that replaces the
particle coordinates $\br_i$ by collective coordinates
$\bx_i(\mathbf{R}) = \br_i + \mathbf{\zeta_i}(\mathbf{R})$ with
\begin{equation*} 
\mathbf{\zeta_i}(\mathbf{R}) =
\sum_{i \neq j}
\mathbf{r}_{ij} \Theta (L_\mathrm{B}-r_{ij})
\bigg(1-\frac{r_{ij}}{L_\mathrm{B}} \bigg)^3 \sum_{n=0}^\mathrm{N_B}
\eta_{s_i s_j,n}
r_{ij}^n,
\end{equation*} 
where $s_i,s_j$ are magnetic quantum numbers of particles $i$ and $j$,
$N_\mathrm{B}=5$ the order of the polynomial, and $L_\mathrm{B}$ is an
optimizable cutoff length. The optimizable coefficients
$\eta_{\alpha\beta,i}$ have to obey the symmetry requirements
$\eta_{\uparrow\uparrow,i} = \eta_{\downarrow\downarrow,i}$ and
$\eta_{\uparrow\downarrow,i} = \eta_{\downarrow\uparrow,i}$. We find
backflow corrections between particles of equal spin to be
insignificant and therefore set $\eta_{\uparrow\uparrow,i} =
\eta_{\downarrow\downarrow,i} = 0$.

The Slater determinant is multiplied by a Jastrow factor $\expe^J$, to
capture the short-distance behavior of the pairwise interaction
potential. We use
\begin{align*}
J=  \sum_{i \neq j}
\Theta (L_\mathrm{J} - r_{ij})
\bigg(1- \frac{r_{ij}}{L_\mathrm{J}} \bigg)^3 
 \sum_{n=0}^{N_\mathrm{J}}
  u_{s_i s_j,n} r_{ij}^n,
\label{eq:jastrow}
\end{align*}
where $N_\mathrm{J}=8$ is the order of the polynomial and
$L_\mathrm{J}$ is an optimizable cutoff length that we choose in
accordance with the cutoff radius of the pseudopotential
\cite{Drummond2004}. The optimizable coefficients $u_{\alpha\beta,i}$
have to obey the symmetry requirements $u_{\uparrow\uparrow,i} =
u_{\downarrow\downarrow,i}$ and $u_{\uparrow\downarrow,i} =
u_{\downarrow\uparrow,i}$, and $u_{\alpha\beta,1}$ is fixed by
requiring zero gradient at the origin. Similar results are obtained
with a Jastrow factor optimized for periodic systems
\cite{Whitehead2016c}.

In the zero-range limit $\kF \Reff = 0$ the Slater-Jastrow trial wave
function captures 93\% of the correlation energy, defined as the
difference in ground state energy between the Hartree-Fock and DMC
results. The backflow transformation captures another 3.5\%, raising
the total to 96.5\%. Backflow transformations are especially important
for negative effective range, where the amount of correlation energy
captured without backflow transformations is only 85\% at
$\kF \Reff = -2$, while a trial wave function with backflow
transformations captures 92\% of the correlation energy.

Observables other than the ground state energy are computed using the
extrapolated estimator $\avg{\hat{A}} = 2 \avg{\hat{A}}_\mathrm{DMC} -
\avg{\hat{A}}_\mathrm{VMC}$, which combines the DMC and VMC
expectation values of the operator $\hat{A}$ to reduce the
bias from linear to quadratic in the difference between the
VMC and DMC wave functions \cite{Ceperley1986}. In
agreement with \refs{Morris2010, Gandolfi2011,Li2011} we find the
results of this extrapolation to be within the statistical error bar
of the DMC estimate and
therefore expect residual errors to be small. We extrapolate to zero
DMC time step and infinite number of walkers to obtain accurate ground
state energies following the procedure detailed in
\secref{timesteps}. We expect that the use of a quadratic DMC
algorithm would give similar results \cite{Mella2000,Sarsa2002}. We
calculate the ground state wave function in the thermodynamic limit
using datapoints for systems with \{66, 114, 162, 186, 294\}
particles; technical details of the extrapolation to infinite system
size are provided in \secref{finitesize}. For the smallest system, we
use $N_\mathrm{PW} = 5$ plane-waves to accommodate the $2 \times 33$
spin-up and down particles, while for the largest system we use
$N_\mathrm{PW} = 10$. We set $N_\mathrm{P} = 6$, allowing us to
accurately describe particles in the virtual bound state for negative
effective range. In total our trial wave function includes 34-39
parameters that we optimize using VMC, before using the trial wave
function as a starting point for our DMC calculations.

\subsection{Ground state energy}
\label{energy}

\begin{figure}
\includegraphics[width=\linewidth]{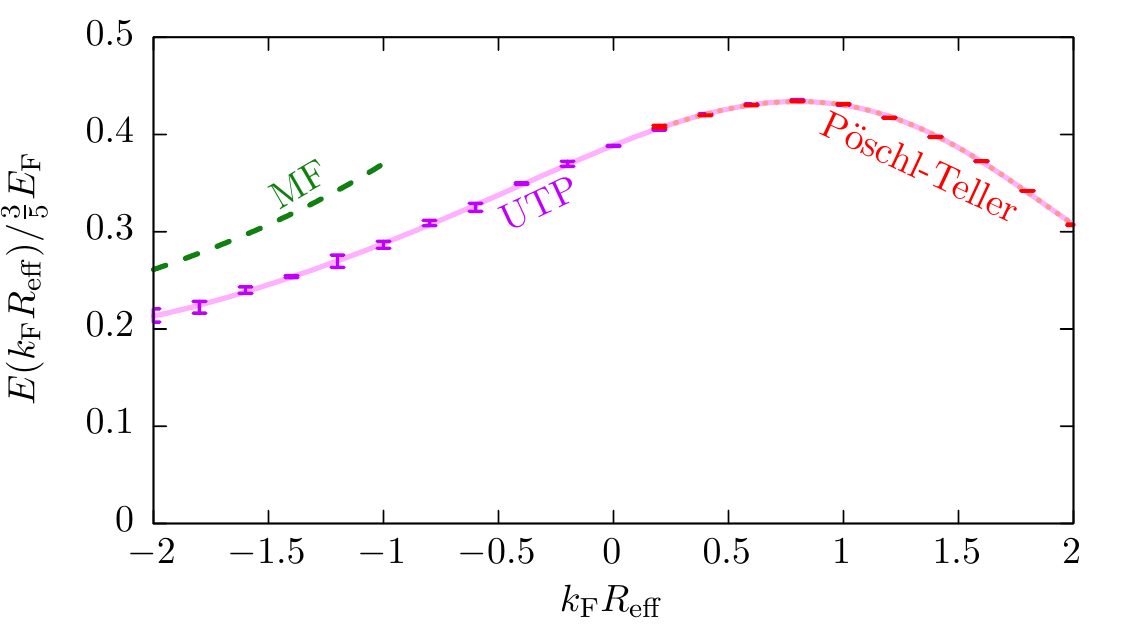}
\caption{(Color online) Ground state energy per particle of the
unitary Fermi gas as a fraction of that of a noninteracting gas with
effective range $\kF \Reff$. Results
obtained using the UTP are shown in purple, and results obtained using
the P\"oschl-Teller interaction for positive effective range in
red. For large negative effective range the mean-field (MF) theory of
\reference{Gurarie2007} is indicated by the dotted green line.}
\label{fig:energy}
\end{figure}

Having developed a pseudopotential that smoothly connects positive and
negative effective ranges and outlined our trial wave function, we are
well positioned to study the ground state properties of resonant Fermi
gases with effective ranges $-2 \leq \kF \Reff \leq 2$. We first study
the ground state energy per particle $E$, plotted as a fraction of
that of a noninteracting gas $E_0 = \frac{3}{5} \EF$, with $\EF$ the
Fermi energy, in \figref{fig:energy}. Starting from the zero-range
case $\kF \Reff=0$, we discuss the large negative and positive
effective range limits. The zero-range case itself will be discussed
in the next section.

We observe a decreasing energy as the effective range tends to $\kF
\Reff=-2$. The potential barrier we saw in \figref{fig:potentials}
rises and decouples the virtual bound sate inside the barrier from the
Fermi sea. This reduces the energy of a pair of opposite spin
particles in the bound state towards the zero energy of the bare
virtual bound state, causing more particles at the Fermi surface to
pair and the energy to approach zero. This behavior is qualitatively the
same as that from the BCS mean-field calculation of
\reference{Gurarie2007}, while quantitatively our DMC energy
approaches their mean-field energy, which is exact only in the limit
$\kF \Reff \to -\infty$.

For positive effective ranges, we observe a maximum value in the
ground state energy $E = 0.432(1) \, E_0$ at $\kF \Reff =
0.8$. For larger effective ranges, the physical range of the
interaction increases so that one particle can now interact
simultaneously with several opposite spin particles, causing the energy
to fall. The rapid decrease in energy in the $\kF \Reff \to \infty$
limit gives rises to a thermodynamic instability that will be
discussed in \secref{stability}.

We also calculate the ground state energy using the alternative
P\"oschl-Teller interaction available for positive effective
ranges. The results for the UTP and P\"oschl-Teller interaction
coincide, demonstrating the universality of the many-body ground state
energy for potentials with equivalent scattering properties in the
Fermi sea.

\subsection{Zero-range limit}

\begin{figure}
\includegraphics[width=\linewidth]{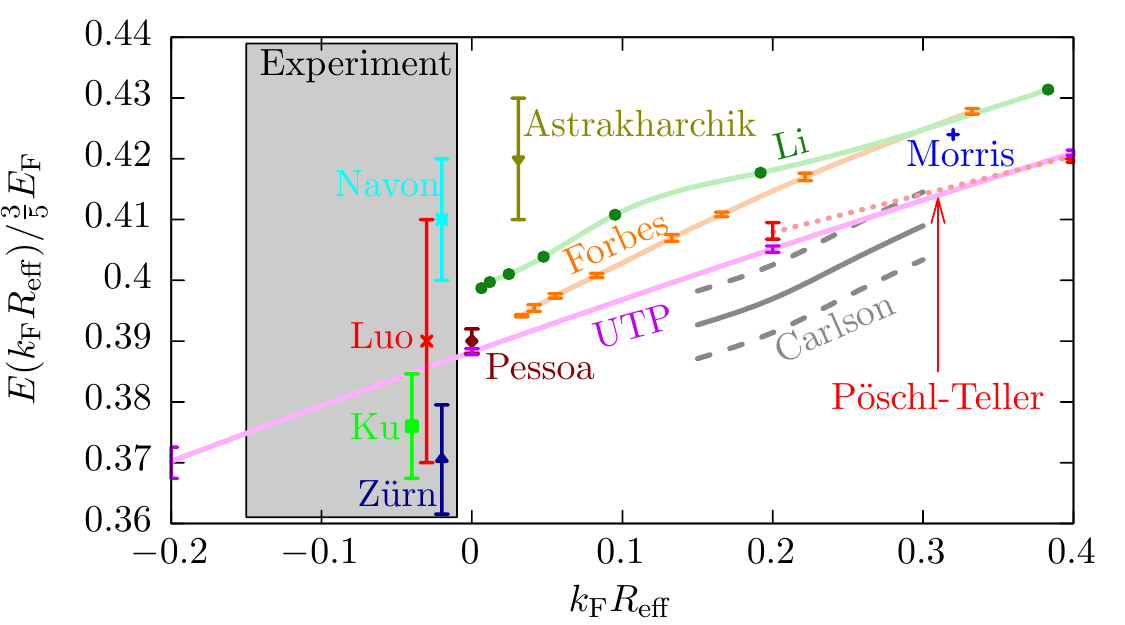}
\caption{(Color online) Ground state energy per particle of the
unitary Fermi gas as a fraction of that of a noninteracting gas with
effective range $\kF \Reff$. The UTP shown in purple and
P\"oschl-Teller in red are results from this work. Also shown are
other numerical results from
\refs{Astrakharchik2004,Morris2010,Carlson2011,Forbes2012,Li2011,Pessoa2015},
where for \reference{Carlson2011} we extrapolated their results to
infinite system size as shown by the solid gray line, with the
uncertainty in our extrapolation indicated by the dashed
lines. Experimental results at $\kF \Reff = 0$ from
\refs{Luo2009,Navon2010,Zurn2013,Ku2012} are shown in the box on the
left, slightly offset from unitarity for improved readability.}
\label{fig:zerorange}
\end{figure}

Having studied the variation of the ground state energy over the full
extent of effective ranges we now focus on the behavior near the zero
range limit $\kF \Reff = 0$ in \figref{fig:zerorange}. For small
effective range the ground state energy per particle can be parameterized as
\cite{Werner2012}
\begin{equation*}
  \frac{E}{\frac{3}{5}\EF} = \xi + \zeta \kF \Reff + \bigo((\kF \Reff)^2),
\end{equation*}
where the Bertsch parameter $\xi$ and $\zeta$ are universal constants
for Galilean invariant continuous space models. From
\reference{VonKeyserlingk2013} we note that the effective coupling is
stronger for more negative effective range and we therefore expect
$\zeta > 0$. We report $\xi = 0.388(1)$, which agrees with the
experimental result of \citet{Luo2009}. Our result is two times the
experimental standard error lower than the result of \citet{Navon2010}, while
it is approximately two standard errors higher than the results of
\citet{Ku2012} and \citet{Zurn2013}. Our statistical error estimates
are negligible in comparison with the experiments, but the fixed node
constraint on the variational wave function introduces a systematic
error, which could explain why our value is higher than the
experimental measurements of \refs{Ku2012,Zurn2013}. Our result agrees
with $\xi = 0.390(2)$ from a DMC calculation by Pessoa et
al. \cite{Pessoa2015}. For the slope we find $\zeta = 0.087(1)$, in
agreement with the auxiliary field result $\zeta = 0.11(3)$ from
\citet{Carlson2011}, but in disagreement with the DMC
result $\zeta = 0.127(4)$ of \citet{Forbes2012}. Before
discussing how this deviating value may be understood as the result of
the computational method employed, we first illuminate how the choice
of pseudopotential influences the results.

To compare the UTP with the P\"oschl-Teller interaction used in
\refs{Morris2010,Forbes2012,Li2011} we calculate the ground state
energy for both potentials using the same trial wave function. As we
have seen before the equivalent phase shift of the UTP and
P\"oschl-Teller interaction guarantees the same ground state energy
for both potentials, but as the effective range is reduced the DMC
energy calculated using the P\"oschl-Teller interaction overestimates
that of the UTP and the error bars for the P\"oschl-Teller interaction
become larger. As QMC is a variational method, it is important to use
an accurate trial wave function. This is especially true for
attractive interactions where the BCS instability requires that the
nodal surface is optimized during a VMC calculation, before fixing the
nodes and further reducing the energy using DMC
\cite{Carlson2003}. The quality of the trial wave function is
described by the variance of the local energy $E_\mathrm{L} =
\Psi^{-1}\hat{H}\Psi$, which is zero for the true ground
state. Because the depth of the of the P\"oschl-Teller interaction
diverges, the local energy variance for $\kF \Reff \leq 0.2$
calculated with the P\"oschl-Teller interaction is more than four
times that calculated with the UTP, explaining the overestimate of the
ground state energy and larger error bars, and confirming the
numerical advantage of our wide and smooth UTP.

Having understood how the smooth UTP results in a lower variational
estimate of the energy, we now compare our results with other DMC
studies
\cite{Astrakharchik2004,Morris2010,Forbes2012,Li2011,Pessoa2015}. We
find a lower variational energy because, besides the smooth UTP, our
study employs a trial wave function that includes more variational
freedom over previous studies. In particular, we have combined the
flexible Jastrow factor, polynomial pairing orbitals and backflow
transformation of \reference{Morris2010} with the plane-wave orbitals
used in \refs{Carlson2003,Forbes2012}. This could explain why
our reported value for the slope $\zeta = 0.087(1)$ is lower than
$\zeta = 0.127(4)$ from \citet{Forbes2012}.

We have also compared our DMC with the auxiliary field QMC study that
is free from the sign problem for a spin balanced system with
attractive interactions and therefore does not require the fixed node
approximation \cite{Carlson2011}. The results in their Fig. 2 display
finite size effects, leading to uncertainty in our extrapolation of
their results to infinite system size. Nevertheless, the extrapolated
ground state energy for effective range $0.15 < \kF \Reff < 0.3$
agrees with our result within $0.01 \, E_0$.

\subsection{Condensate fraction}

\begin{figure}
\includegraphics[width=\linewidth]{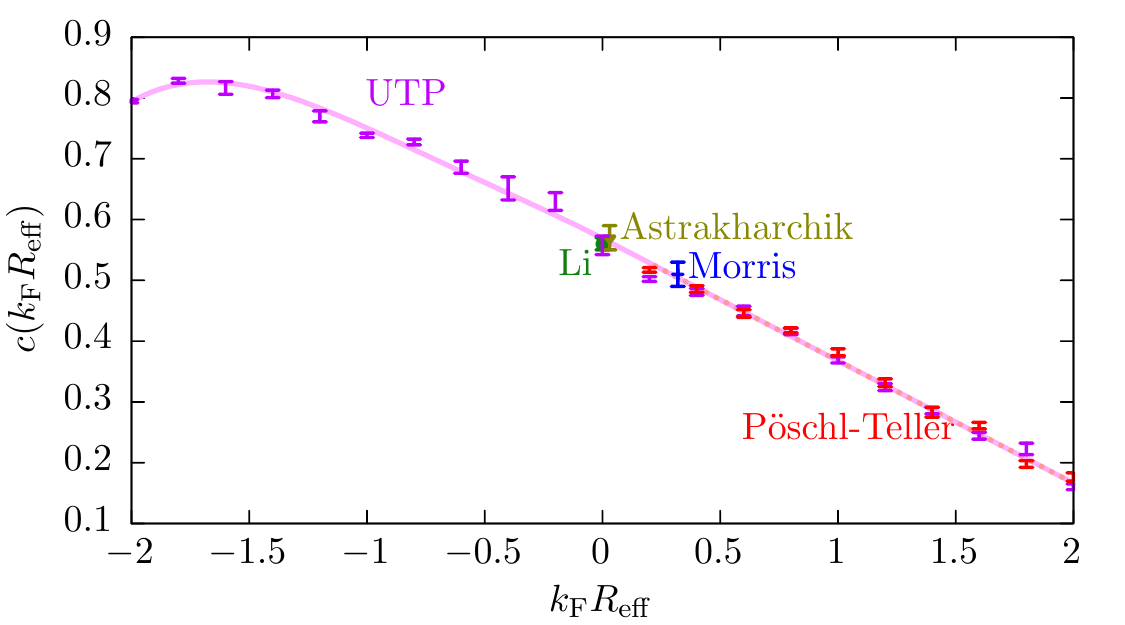}
\caption{(Color online) Condensate fraction as a function of effective
  range calculated using the UTP in purple and  P\"oschl-Teller
  interaction in red for comparison. We also plot the
  results obtained near the zero-range limit 
  by \refs{Astrakharchik2005,Morris2010,Li2011}.}
\label{fig:condfrac}
\end{figure}

Having studied the variation of the ground state energy we now examine
other expectation values starting with the condensate fraction. A
defining feature of a superconductor is the existence of a condensate
that introduces correlations between Cooper pairs of opposite spin particles
irrespective of their separation. Correlations between pairs of
opposite spins are naturally captured by the off-diagonal two-body
density matrix
\begin{equation*}
  \label{eq:2bdm} \rho_{\downarrow \uparrow}^{(2)}(\br_1', \br_2';
\br_1, \br_2) = \langle c_\uparrow^\dagger(\br_1')
c_\downarrow^\dagger(\br_2') c_\downarrow(\br_2)
c_\uparrow(\br_1) \rangle,
\end{equation*}
where $c_\alpha^{\dagger}(\br)$ is the fermionic creation and
$c_\alpha(\br)$ the annihilation operator for a particle with spin
$\alpha$ at position $\br$. It is convenient to work in coordinates
that make the separation between two pairs, $\bR =
\frac{1}{2}(\br_1'+\br_2')-\frac{1}{2}(\br_1+\br_2)$, and the size of
the pairs $\br = \br_1 - \br_2$, and $\br' = \br'_1-\br'_2$
explicit. In the limit $R = |\bR| \to \infty$ the two-body density
matrix is proportional to the condensate fraction
$c$ \cite{Leggett2006,Astrakharchik2005}
\begin{equation}
  \label{eq:condfrac}
  \rho_{\downarrow
    \uparrow}^{(2)}\bigg(\bR+\frac{\br'}{2},\bR-\frac{\br'}{2};
  \frac{\br}{2},-\frac{\br}{2}\bigg)
  \to c \frac{N}{2}
  \phi^*(|\br'|)\phi(|\br|),
\end{equation}
where $\phi(r)$ is the complex pair wave function normalized to
reciprocal volume $1/\Omega$, and $N$ is the number of particles. In
the normal phase where correlations between pairs vanish as $R \to
\infty$, $c=0$, while for a superconducting phase the pairs remain
correlated however far apart they are and the fraction of particles in
the condensate is $0<c\leq 1$.

The numerical computation of the condensate fraction using the
relation above is complicated as it requires an extrapolation to the
$R \to \infty$ limit \cite{Astrakharchik2005,Morris2010,Li2011}. This
motivates us to use a Fourier transform to capture the long-distance
behavior as a zero-momentum mode, in order to accurately compute the
condensate fraction using all accumulated samples of the two-body
density matrix across the entire simulation cell, following the
procedure outlined in \appref{app:condensate}.

The condensate fractions calculated with the UTP and P\"oschl-Teller
interactions agree as seen in \figref{fig:condfrac}, confirming the
accuracy of the UTP.  We also show data from references
\cite{Astrakharchik2005,Morris2010,Li2011} for comparison and, after
taking into account the effective ranges used, observe good agreement
between results. In the zero-range limit $\kF \Reff = 0$, we report $c
= 0.56(2)$ and the negative slope for the condensate fraction is
consistent with the positive slope for the energy encountered earlier,
as the breaking of Cooper pairs increases the energy. There is a
maximum in the condensate fraction at $\kF \Reff \approx -1.8$ of
$0.83(1)$. For more negative effective range, the virtual bound states
decouple from the Fermi sea, and so, although the particles remain
paired, they no longer interact with each other and correlations
between pairs vanish, causing the condensate fraction to decrease
\cite{Gurarie2007}.

\subsection{Momentum distribution}

\begin{figure}
\includegraphics[width=\linewidth]{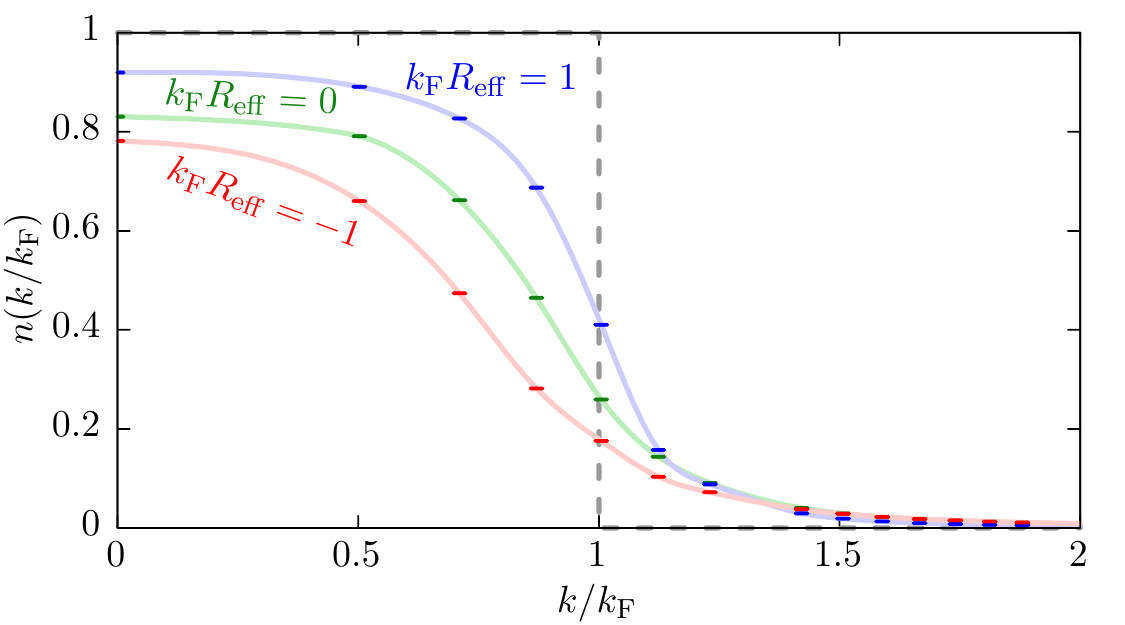}
\includegraphics[width=\linewidth]{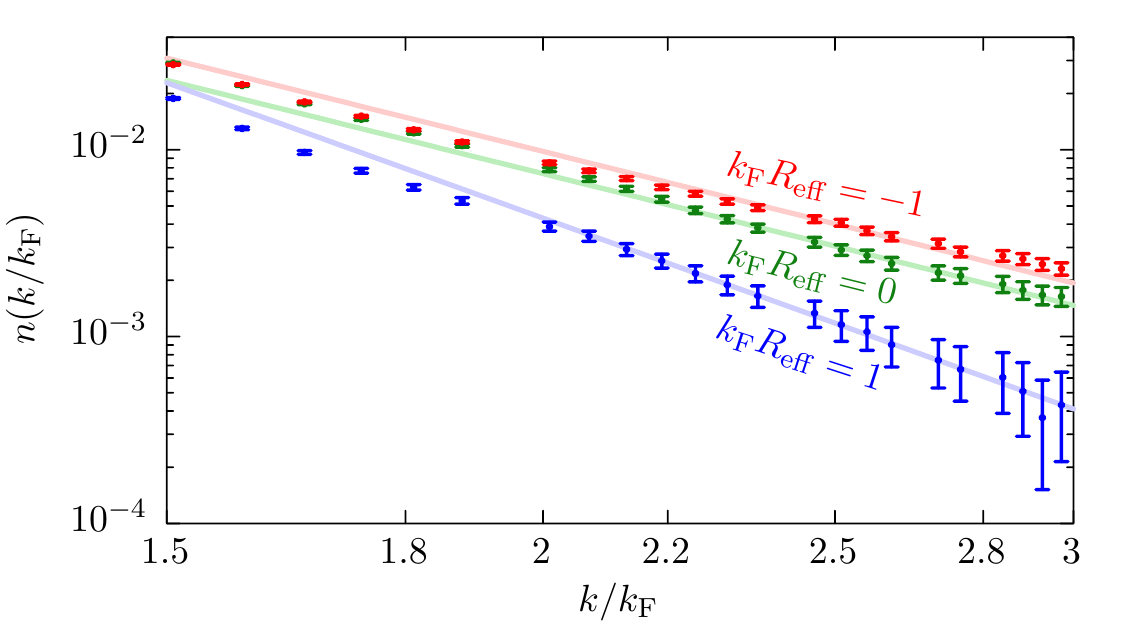}
\caption{(Color online) (Top) Momentum distribution $n(k)$ for $\kF
\Reff = \{-1,0,1\}$. The noninteracting distribution is indicated by
the gray dashed line. (Bottom) Tail of the momentum distribution on
logarithmic axes. The lines indicate a weighted least squares fit of
$n(k)=C/k^\alpha$ to the momentum tail $k > 2.2 \kF$.}
\label{fig:momdist}
\end{figure}

Having surveyed how the ground state energy and condensate fraction
vary with effective range, we now select three characteristic
effective ranges $\kF \Reff = \{-1, 0, 1\}$ to study one- and two-body
correlation functions. In this section we study the momentum
distribution shown in \figref{fig:momdist}. In the limit $\Reff \to \infty$ the
physical range of the potential diverges, approaching a constant
background potential and so the momentum distribution approaches
that of a noninteracting system. When the effective range is
decreased from positive to negative, the sharp cutoff at the Fermi
momentum disappears as weight is moved from low momenta to the high
momentum tail characteristic of a state of paired particles. 

The tail of the momentum distribution at unitarity in the zero-range
limit is $n(k) \to C/k^4$, where $C$ is Tan's contact
\cite{Tan2008,*Tan2008a,*Tan2008b}. As shown by
\reference{Braaten2008} this result extends to $\Reff < 0$ and the
contact becomes a function of effective range $C(\kF \Reff)$. In the
zero-range limit we report $C(0)/\kF^4 = 0.119(1)$, which is in
reasonable agreement with 0.1147(3) \cite{Gandolfi2011} and 0.0961(1)
\cite{Pessoa2015} computed using different trial wave functions. Our
results for $\kF \Reff = -1$ are consistent with a $\sim 1/k^4$ tail,
and we find an increased value for the contact $C(-1)/\kF^4 =
0.157(3)$ as expected when more particles pair.  For positive
effective ranges we observe a more rapidly decaying tail.

\subsection{Pair-correlation function}

\begin{figure}
\includegraphics[width=\linewidth]{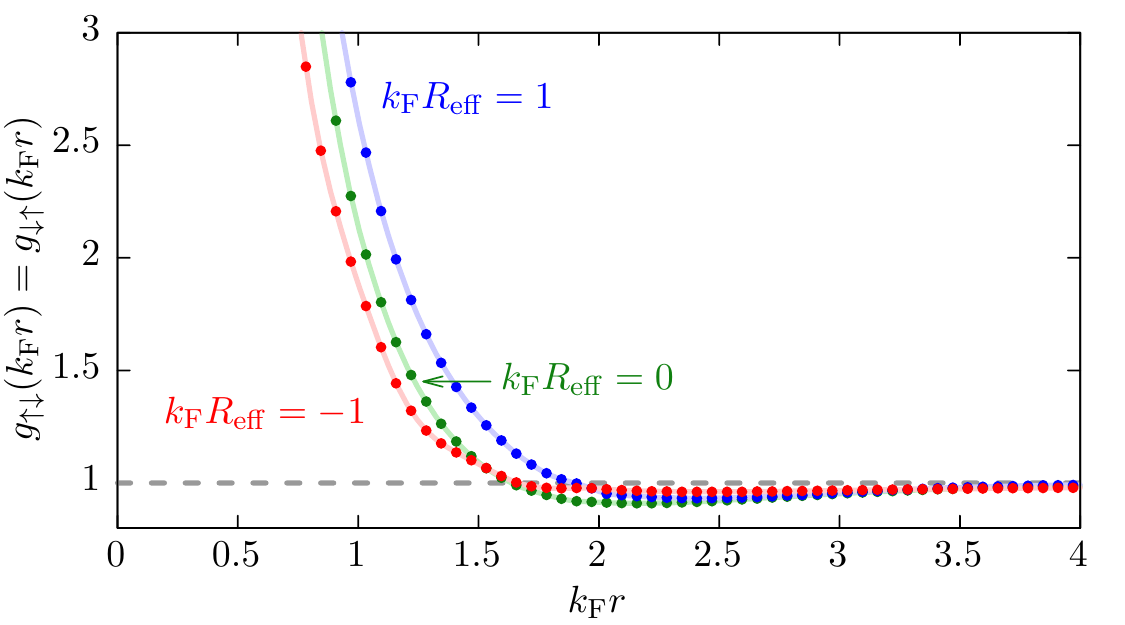}
\includegraphics[width=\linewidth]{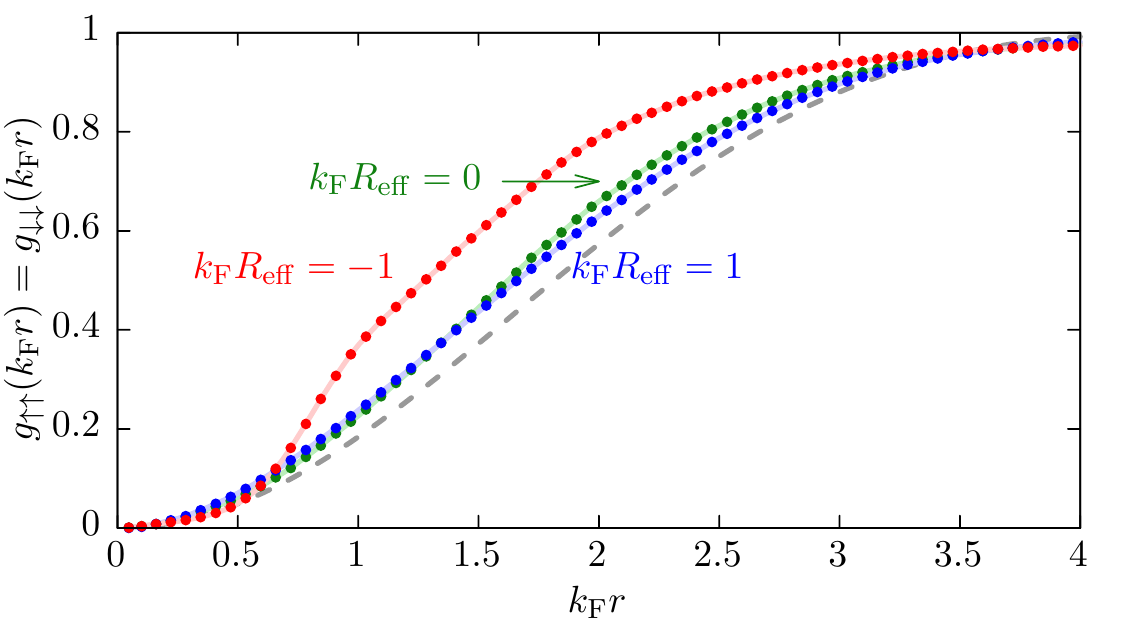}
\caption{(Color online) Pair-correlation function for opposite (top) and
equal (bottom) spins for $\kF \Reff = \{-1,0,1\}$. The noninteracting
correlation function is indicated by the gray dashed line.}
\label{fig:paircorr}
\end{figure}

To better understand the two-body interactions that cause the
deformation of the Fermi surface, we show the pair-correlation
function for opposite and equal spins in \figref{fig:paircorr}. For
opposite spins, we correct the pair-correlation function for short-range effects
due to the particular form of the pseudopotential \cite{Lloyd-Williams2015}
\begin{equation*}
  g_{\uparrow \downarrow}(r) = 
  \frac{g_{\uparrow\downarrow}^{\mathrm{2-body, exact}}(r)}{g_{\uparrow
      \downarrow}^{\mathrm{2-body, pseudo}}(r)}
  g_{\uparrow \downarrow}(r),
\end{equation*}
where $g_{\uparrow\downarrow}^{\mathrm{2-body,\{exact,pseudo\}}}(r)$ are the
pair-correlation functions for the two-body problem computed using the
exact and pseudopotential wave functions respectively. Since our UTP is
norm-conserving \cite{Bugnion2014,Whitehead2016a} no correction is
necessary outside of the interaction region.

For opposite spins, the pair-correlation function naturally shows that
due to the attractive interaction the particles are more likely to be
found in close proximity compared to the noninteracting case. In the
zero-range limit the pair correlation diverges at short inter-particle
distances as $\sim 1/r^2$ \cite{Tan2008,*Tan2008a,*Tan2008b}. This
divergence becomes stronger for negative effective range where
particles of opposite spins are more likely to be found in pairs,
whereas for positive effective range the particles are further apart
compared to the zero-range case. The dominant contribution to the
correlations between equal spins at positive and zero effective range
is the exchange-correlation hole due to the Pauli exclusion
principle. The volume of the exchange-correlation hole diminishes as
the effective range becomes negative, because the fermions are more
likely to be paired in the virtual bound state and behave as composite
bosons.

\section{Thermodynamic stability}
\label{stability}

We saw in \figref{fig:energy} that the ground state energy of a Fermi
gas at unitarity falls rapidly with increasing effective range. This
raises the possibility of a thermodynamic instability towards phase
separation into a high density phase with a high value of $\kF \Reff$,
and so a large negative energy, and a low density phase. To
analyze this possibility we first assess the behavior of the ground state
energy using a mean-field approximation before investigating the
thermodynamic instability.

\subsection{Hartree-Fock theory}
The proposed collapse into the dense phase means the dimensionless
physical interaction range $\kF \Reff$ diverges and the interaction
potential approaches a constant background potential. In this limit
the wave function approaches that of a noninteracting system so we can
use the Hartree-Fock approximation to estimate the ground state energy
per particle as
\begin{align}
  \label{eq:hartree-fock}
  E_{\mathrm{HF}} &= \frac{3 \EF}{5} + \frac{1}{4} \int \du^3r \, V(\br) n(\br)
  \nonumber \\
  &=  \frac{3 \EF}{5} \bigg(1-\frac{10 \pi}{108} \kF \Reff \bigg),
\end{align}
with $n(\br)$ the density, which in our case is uniform so $n(r) = n$,
and the factor of $1/4$ accounts for the fact that interactions act only
between particles of opposite spin. The dependence on the explicit
form of the interaction potential only enters via the integral, and
the result is independent of the choice for the potential-well,
P\"oschl-Teller, or UTP.

\subsection{Stability}

To assess the thermodynamic stability we consider the Helmholtz free
energy density $F/\Omega$, with $\Omega$ the volume. The Helmholtz
free energy is $F = E - TS$, with temperature $T=0$ and $S$ the
entropy. For a thermodynamically (meta)stable phase the free energy
density is required to be a convex function of density, so
$d^2F/dn^2>0$, whereas if $d^2F/dn^2<0$ the system phase separates
\cite{Landau2001}.

\begin{figure}
\includegraphics[width=\linewidth]{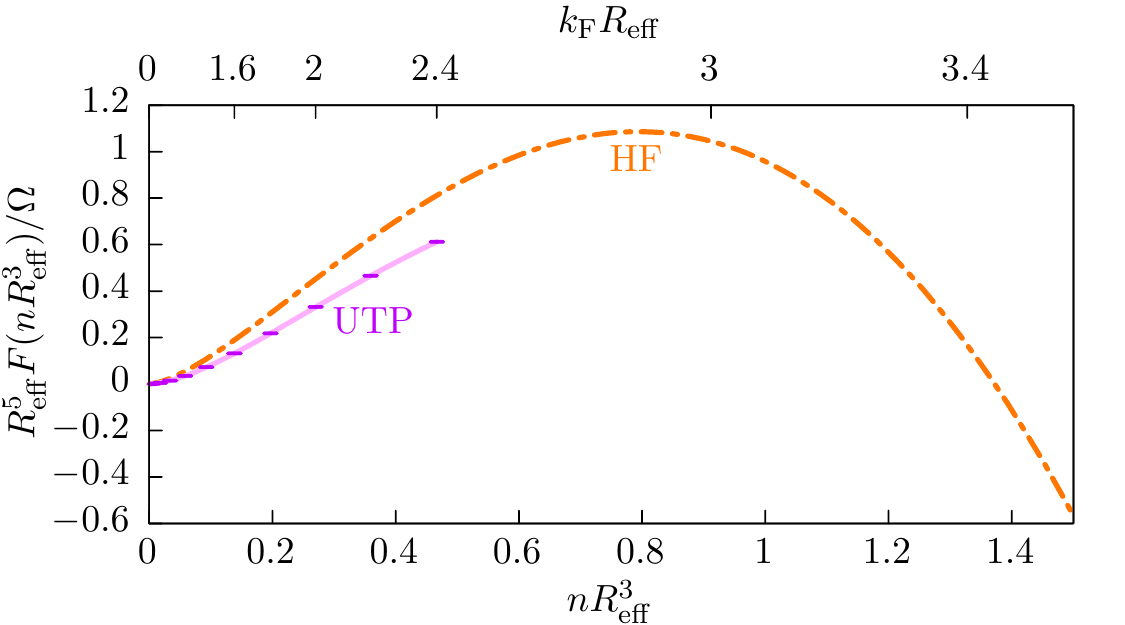}
\includegraphics[width=\linewidth]{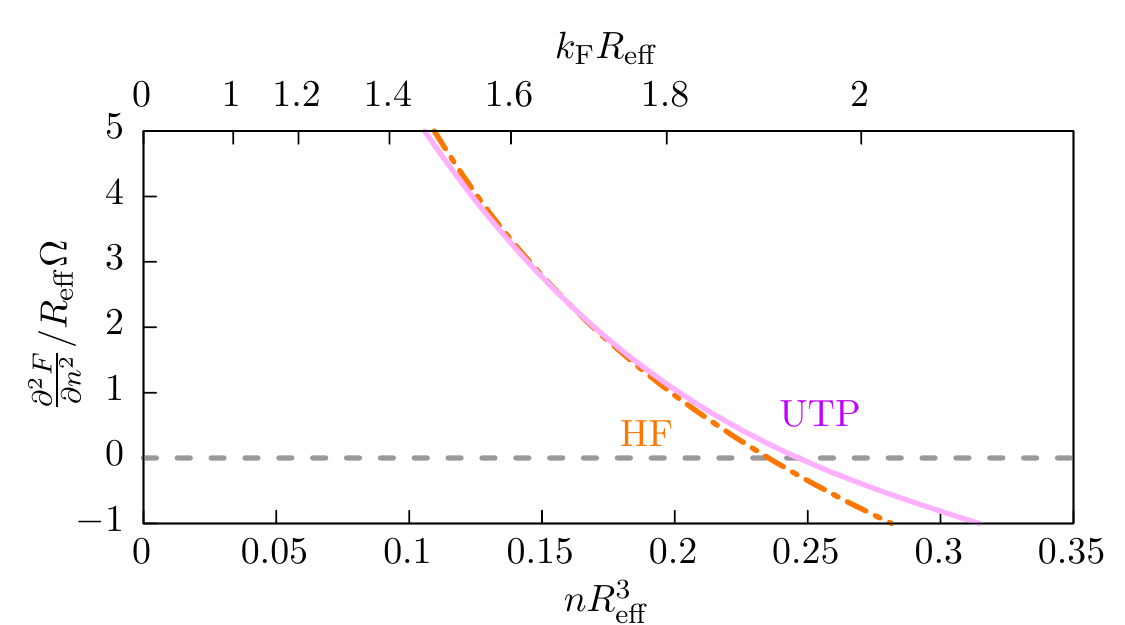}
\caption{(Color online) (Top) Energy density as a function of the
Fermi gas density on the bottom axes and Fermi wave vector on the top
axes. HF is the Hartree-Fock theory
\eqnref{eq:hartree-fock} shown by the green dashed line. We also show
our DMC results obtained using the UTP in purple, fitted with cubic
splines. (Bottom) Curvature for the Hartree-Fock theory and UTP. The
spinodal point estimated from UTP data at $n \Reff^3 \approx 0.25$
($\kF \Reff \approx 1.9$) is near the spinodal point calculated using
the Hartree-Fock theory at $n \Reff^3 = 0.235$ ($\kF \Reff = 1.91$).}
\label{fig:stability}
\end{figure}

In \figref{fig:stability} we show the free energy density derived from
\eqnref{eq:hartree-fock} as a function of the Fermi gas density, as
well as the curvature derived from this free energy density. A
spinodal point where the free energy density turns from convex to
concave occurs at $n \Reff^3 = 0.235$. This signals the onset of phase
separation into an infinitely dense phase and a low density phase for
effective range $n \Reff^3 > 0.235$, that is $\kF \Reff > 1.91$.

In the same figure we also show our DMC data obtained using the
UTP. Our DMC simulations are insensitive to phase separation as the
trial wave function has insufficient overlap with that of a phase
separated state, so we can address the full extent of effective
ranges. We interpolate our DMC results with cubic splines to determine
$d^2F/dn^2$, which show that the spinodal point is at 
$n \Reff^3 \approx 0.25$, $\kF \Reff \approx 1.9$, consistent with the
Hartree-Fock result.

We establish that resonant gases with effective range $\kF \Reff
\gtrsim 1.9$ are thermodynamically unstable to particle collapse. The
instability of a Fermi gas with finite range interactions at unitarity
is reminiscent of the instability for neutron matter at finite
scattering lengths, where a three-body repulsive force is necessary to
ensure thermodynamic stability \cite{Friedman1981}. The concerns of
\reference{Forbes2012} that the P\"oschl-Teller interaction might harbor a
many-body bound state is a precursor to the instability presented here.

\section{Conclusion}

We have proposed the UTP as an interaction potential to model resonant
scattering of fermions with varying effective interaction
range. Unlike other potentials, the UTP smoothly connects the positive
and negative effective range regimes. Moreover, at the midpoint
between those regimes where the effective range is zero, the UTP
remains smooth, extended in space, and of finite depth. This allows us to perform an
accurate calculation of the ground state properties as we can directly
simulate the zero-range limit, with no need for extrapolations.

Exploiting the numerical advantages of the UTP and an improved
estimator for the condensate fraction, we have performed DMC ground
state calculations for resonant gases as a function of effective
interaction range. In the zero-range limit, we report values for the
universal constants of $\xi = 0.388(1)$ and $\zeta = 0.087(1)$, for
the contact $C/\kF^4 = 0.119(1)$, and for the condensate fraction $c =
0.56(2)$. Furthermore, by studying the momentum distribution and pair
correlation functions, we have demonstrated how the system evolves
from a state of independent pairs of opposite spin particles for
negative effective range, to the strongly interacting state in the
zero-range limit $\kF \Reff = 0$, and finally to the weakly
interacting BCS superconductor for positive effective range.  We find
resonant gases with effective range $\kF \Reff \gtrsim 1.9$ are
unstable to phase separation into an infinitely dense phase and a
vacuum phase containing no particles.

Having covered the complete gamut of effective interaction ranges, we
expect our results will be relevant for cold atom gases with both
broad and narrow Feshbach resonances. On the positive effective range
side we also expect our results to be relevant for neutron
matter. Furthermore, the UTP formalism, extended here to include
the effective range term, will be useful for future studies of both
contact and finite ranged interactions.

\acknowledgments {The authors thank Pablo L\'opez R\'ios, Thomas
Whitehead, Neil Drummond, Richard Needs, Stefano Giorgini, and Jordi Boronat for
useful discussions. The authors acknowledge the financial support of
the EPSRC [EP/J017639/1]; LMS acknowledges financial support from the
Cambridge European Trust, VSB Fonds, and the Prins Bernhard Cultuurfonds;
and GJC acknowledges the financial support of the Royal Society and
Gonville \& Caius College.  Computational facilities were provided by
the University of Cambridge High Performance Computing Service. There
is Open Access to this paper and data \cite{Schonenberg2016a}.}

\appendix

\section{Details of QMC extrapolations}
In this section we provide technical details of the extrapolations
employed to acquire accurate QMC data. To accurately extract the
ground state energy it is important to extrapolate to zero time step
and infinite walker population, discussed in the next section, and to
the thermodynamic limit, discussed in the second section.

\subsection{Time step and walker population extrapolation}
\label{timesteps}

In DMC the imaginary time evolution operator $\expe^{-\hat{H}
\Delta\tau}$ is applied at each time step $\Delta \tau$ to a trial
wave function, represented by a finite number of walkers, to project out
the ground state \cite{Foulkes2001}. The gradient of the energy as a
function of time step is expected to be proportional to the local
energy variance \cite{Lloyd-Williams2015,Whitehead2016} and the true ground state is
recovered by extrapolating to zero time step and infinite walker
population \cite{Lee2011}.

\begin{figure}
\includegraphics[width=\linewidth]{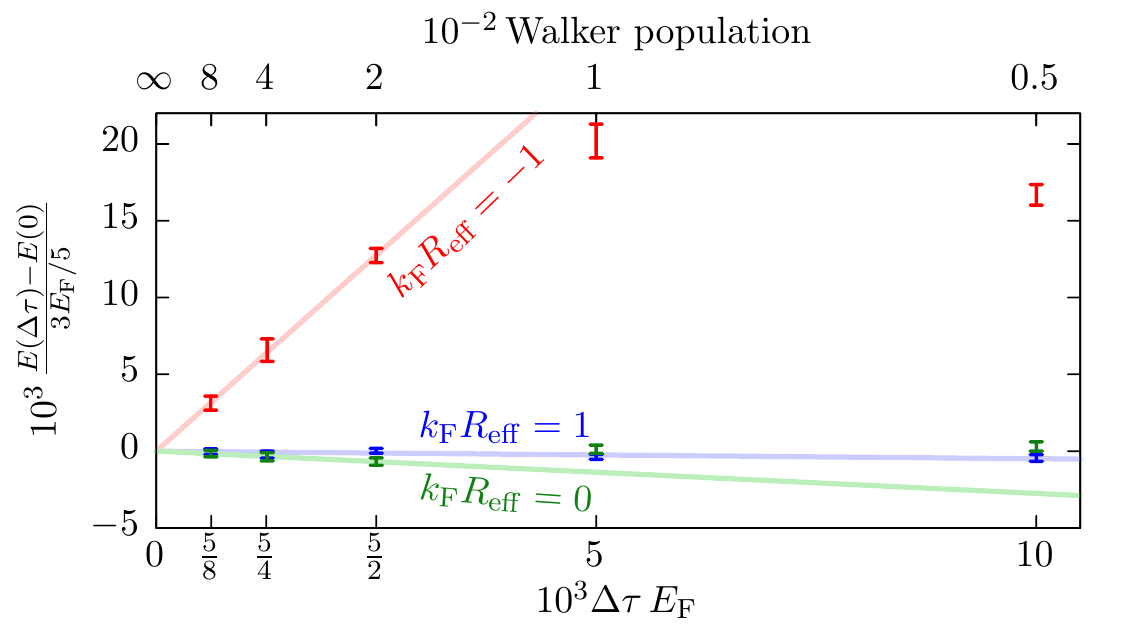}
\caption{(Color online) Ground state energy for $\kF \Reff = \{-1, 0,
1\}$ as a function of time step (bottom axes) and walker population
(top axes). The straight lines show a weighted least squares fit to
the data points in the linear regime.}
\label{fig:timestep}
\end{figure}

In practice we perform these extrapolations simultaneously by reducing
the time step by a factor of two, while increasing the walker
population by the same factor, as in \figref{fig:timestep}. For
optimal efficiency the computational effort should be increased by
$2\sqrt{2}$ for each division of the time step by 2
\cite{Lee2011}. For effective range $\Reff \geq 0$ our trial wave
function optimized using VMC results in small local energy
variances. The variation with energy and walker population is
therefore small, less than $10^{-3} E_0$ in the linear regime, which
for $\kF \Reff = 1$ extends to time steps $\Delta \tau \EF \leq
10^{-2}$ and for $\kF \Reff = 0$ to time steps $\Delta \tau
\EF \leq 2.5 \mytimes 10^{-3}$. However, for negative effective range
$\Reff<0$ the local energy variance is larger and we observe a
significant variation with time step and walker in the linear regime
extending up to time steps $\Delta \tau \EF \leq 2.5 \mytimes
10^{-3}$. We have performed additional tests to show that the
variation of the energy originates from the reduction in time step and
not from the increase in walker population. Extrapolating to zero time
step is essential as even the smallest time step used $\Delta \tau \EF
= 0.625 \mytimes 10^{-3}$ introduces a systematic error to the ground
state energy of $3 \mytimes 10^{-3} E_0$, and smaller time steps would
require a large number of steps to exceed the auto-correlation time of
the random Monte Carlo walk. We conclude that extrapolating to zero
time step and infinite walker population is essential to accurately
calculate ground state energies.

\subsection{System size extrapolation}
\label{finitesize}

Having extrapolated to zero time step and infinite walker population,
we now extrapolate to infinite system size. As the length scale
associated with the features of our interaction potential is less than
the average inter-particle separation, we expect system size effects to
be dominated by the kinetic energy term in the Hamiltonian. The
finite-size error in the kinetic term originates from the
discretization of the plane-wave wave vectors, which in three
dimensions is proportional to the reciprocal number of particles $1/N$
\cite{Drummond2008}.

\begin{figure}
\includegraphics[width=\linewidth]{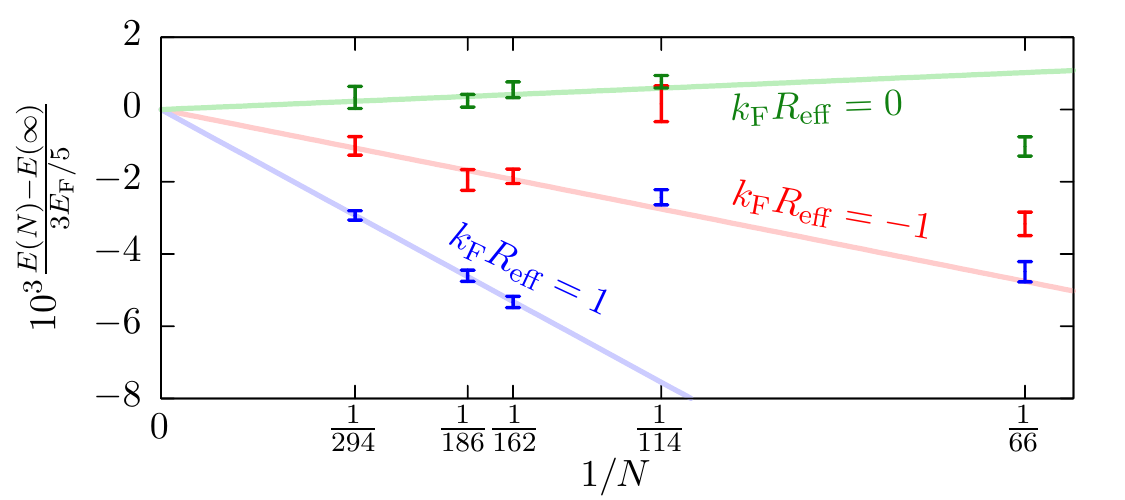}
\caption{(Color online) Ground state energy for $\kF \Reff
  = \{-1, 0, 1\}$ as a function of the number of particles. The
  straight lines show a weighted least squares fit to the data points 
  with $N \geq 162$.}
\label{fig:finitesize}
\end{figure}

Exploiting our smooth pseudopotential to create low-variance trial
wave functions, we can study systems with up to 294 particles. Results
are shown in \figref{fig:finitesize}, where we observe the expected
linear regime for systems with more than 162 particles. For $\Reff
\leq 0$, finite size effects are $< 2 \mytimes 10^{-3} E_0$ for
systems with more than 162 particles, where the particles are bound in
pairs described by the polynomial term in the pairing orbitals. In
contrast, for $\kF \Reff = 1$ the plane-wave term dominates and the
trial wave function is closer to that of a noninteracting system,
thus displaying larger finite-size effects with variations in energy
up to $5 \mytimes 10^{-3} E_0$ as the system size is decreased from an
infinite number of particles to 162 particles. The residual errors in
the extrapolation are $< 10^{-3} E_0$ and we conclude that
extrapolating to infinite system size using systems with at least 162
particles is essential to obtain accurate predictions for the ground
state energy.

\section{Evaluating the condensate fraction}
\label{app:condensate}

A central property of a superconductor is the existence of a
condensate of pairs of particles. The condensate manifests itself as a
macroscopic eigenvalue of the two-body
density matrix for opposite spins irrespective of the distance
between the pairs of opposite spin particles, i.e.  off-diagonal long range order
\cite{Yang1962}. In practice the limiting behavior of the
two-body density matrix is often used to compute the condensate
fraction \cite{Astrakharchik2005,Morris2010,Li2011}, thereby ignoring available
knowledge of the two-body density matrix at short distances and in the
corners of the simulation cell. Here, we propose a Fourier
transform to exploit knowledge of a modified two-body density matrix
over the entire simulation cell to accurately estimate the condensate
fraction. We show that this improved estimator gives direct access to
the condensate fraction.

We consider the BCS wave function \cite{Annett2004}
\begin{equation*}
  \label{eq:bcs-wave function} \ket{\Psi_{\mathrm{BCS}}} = \prod_\bk
(u_\bk^* + v_\bk^* c^\dagger_{\uparrow \bk} c^\dagger_{\downarrow
-\bk}) \ket{0},
\end{equation*} with $u_\bk, v_\bk$ the usual complex coherence
factors to evaluate the expectation values in this section. As
demonstrated by the Eagles-Leggett mean-field theory of the BEC-BCS
crossover this wave function is qualitatively correct even in the
strong coupling limit \cite{Eagles1969,Leggett1980}. We introduce
the order parameter $F(\br) = \avg{c_{\downarrow}(\br)
c_{\uparrow}(\br)}$, related to the pair wave function introduced
earlier as $F(r) = \sqrt{cN/2} \phi(r)$. Expressed in terms of the
coherence factors $F(\br)$ is
\begin{equation*}
 \label{eq:cond-wave function} F(\br) = \frac{1}{\Omega} \sum_{\bk}
\expe^{\ii \bk \cdot \br} u_\bk v_\bk^*,
\end{equation*} 
with $\Omega$ the volume. $F(\br)$ is the eigenfunction of the
off-diagonal two-body density matrix at large inter-pair separation $R$ and the
condensate fraction $c$ is defined in terms of its macroscopic
eigenvalue \cite{Leggett2006}
\begin{align*}
 \label{eq:condensate-fraction} c & =  \frac{2\Omega}{N} \int_\Omega
\du^3r |F(\br)|^2 \nonumber \\ & =  \frac{2}{N} \sum_\bk |u_\bk|^2
|v_\bk|^2.
\end{align*}

To compute the condensate fraction numerically we introduce the
spatially averaged one- and two-body density matrices. We use the projected
two-body density matrix obtained by setting the separation between
particles in each pair equal to each other, $\br = \br'$, thereby eliminating one
coordinate to integrate over \cite{DePalo2002,Ortiz2005}
\begin{align*}
  \bar{\rho}^{(1)}_{\alpha}(\bR)
  =& \frac{1}{\Omega} \int_\Omega
  \du^3\bar{r} \rho^{(1)}_{\alpha}(\bar{\br}+\bR,\bar{\br}), \nonumber \\
  \bar{\rho}_{\downarrow\uparrow}^{(\mathrm{2})}(\bR)
  =& \frac{1}{\Omega^2} \int_\Omega
  \du^3 \bar{r} \, \du^3 r \nonumber \\ 
  & \rho_{\downarrow
    \uparrow}^{(2)}\bigg(\bar{\br}+\bR+\frac{\br}{2},\bar{\br}+\bR-\frac{\br}{2};
  \bar{\br}+\frac{\br}{2},\bar{\br}-\frac{\br}{2}\bigg),
\end{align*}
where $\rho^{(1)}_{\alpha}(\br',\br) = \avg{c_\alpha^\dagger(\br')
  c_\alpha(\br)}$ is the one-body density matrix for spin
$\alpha$.

To remove known short-ranged one-body contributions from the two-body
density matrix we follow \reference{Astrakharchik2005} and introduce an
estimator for the condensate fraction
\begin{equation*}
  \label{eq:condensate-estimator}
  c_{\uparrow \downarrow}(\bR) =  \frac{2 \Omega^{3/2}}{N}
  \big( \bar{\rho}^{(2)}_{\uparrow
    \downarrow}(\bR) -
 \bar{\rho}^{(1)}_{\uparrow}(\bR) \bar{\rho}^{(1)}_{\downarrow}(\bR) \big).
\end{equation*}
Using \eqnref{eq:condfrac} we find at large radius $R$
\begin{equation*}
  \lim_{R \to \infty} c_{\downarrow\uparrow}(R) = c/\sqrt{\Omega}.
\end{equation*}

The extrapolation to the large $R$ limit is problematic in numerical
studies where information is available for finite $R$ values only, and
would neglect information available in the simulation cell at smaller
distances and further out in the corners of the simulation
cell. Instead, we propose a Fourier transform to capture the long
distance behavior of the two-body density matrix as a discontinuity at
small momentum. Defining the Fourier transform pair
as
\begin{align*}
 \label{eq:fourier-transform}
  f(\br) &= \frac{1}{\sqrt{\Omega}} \sum_\bk \expe^{-\ii \bk \cdot \br} f_\bk, \\
  f_\bk &= \frac{1}{\sqrt{\Omega}} \int_\Omega \du^3r \, \expe^{\ii \bk \cdot \br} f(\br),
\end{align*}
we compute the Fourier transform of the modified two-body
density matrix to give us direct access to the condensate
fraction $c$
\begin{align*}
  c_{\uparrow \downarrow \bq} & =  \frac{2 \Omega^{3/2}}{N} \big( \bar{\rho}^{(2)}_{\uparrow
    \downarrow \bq} -  
 \sum_\bk \bar{\rho}^{(1)}_{\uparrow \bk}
 \bar{\rho}^{(1)}_{\downarrow \bq-\bk} \big) \nonumber \\
 & =   \frac{2}{N} \sum_\bk \big[ \avg{c^\dagger_{\uparrow \bk} c^\dagger_{\downarrow \bq - \bk} 
    c_{\downarrow \bq - \bk} c_{\uparrow \bk}} \nonumber \\
 &  \hspace{50pt} - \avg{c^\dagger_{\uparrow \bk} c_{\uparrow \bk}}
   \avg{c^\dagger_{\downarrow \bq-\bk} c_{\downarrow \bq-\bk}} \big] \nonumber \\
 & =   \frac{2 \delta_{\bq \bzero}}{N} \sum_\bk |v_\bk|^2
 |u_\bk|^2 \nonumber \\
 & =  \delta_{\bq \bzero} c,
\end{align*}
with $\delta_{ab}$ the Kronecker delta function. The condensate
fraction exists as a discontinuous peak at zero momentum as expected,
and due to the subtraction of the one-body density matrix the
condensate fraction estimator contains no other contributions.  We
exploit this relation to accurately compute the condensate fraction
using all accumulated samples of the modified two-body density matrix in
the simulation cell.

\bibliographystyle{apsrev4-1}
\bibliography{library}

\begin{thebibliography}{62}%
\makeatletter
\providecommand \@ifxundefined [1]{%
 \@ifx{#1\undefined}
}%
\providecommand \@ifnum [1]{%
 \ifnum #1\expandafter \@firstoftwo
 \else \expandafter \@secondoftwo
 \fi
}%
\providecommand \@ifx [1]{%
 \ifx #1\expandafter \@firstoftwo
 \else \expandafter \@secondoftwo
 \fi
}%
\providecommand \natexlab [1]{#1}%
\providecommand \enquote  [1]{``#1''}%
\providecommand \bibnamefont  [1]{#1}%
\providecommand \bibfnamefont [1]{#1}%
\providecommand \citenamefont [1]{#1}%
\providecommand \href@noop [0]{\@secondoftwo}%
\providecommand \href [0]{\begingroup \@sanitize@url \@href}%
\providecommand \@href[1]{\@@startlink{#1}\@@href}%
\providecommand \@@href[1]{\endgroup#1\@@endlink}%
\providecommand \@sanitize@url [0]{\catcode `\\12\catcode `\$12\catcode
  `\&12\catcode `\#12\catcode `\^12\catcode `\_12\catcode `\%12\relax}%
\providecommand \@@startlink[1]{}%
\providecommand \@@endlink[0]{}%
\providecommand \url  [0]{\begingroup\@sanitize@url \@url }%
\providecommand \@url [1]{\endgroup\@href {#1}{\urlprefix }}%
\providecommand \urlprefix  [0]{URL }%
\providecommand \Eprint [0]{\href }%
\providecommand \doibase [0]{http://dx.doi.org/}%
\providecommand \selectlanguage [0]{\@gobble}%
\providecommand \bibinfo  [0]{\@secondoftwo}%
\providecommand \bibfield  [0]{\@secondoftwo}%
\providecommand \translation [1]{[#1]}%
\providecommand \BibitemOpen [0]{}%
\providecommand \bibitemStop [0]{}%
\providecommand \bibitemNoStop [0]{.\EOS\space}%
\providecommand \EOS [0]{\spacefactor3000\relax}%
\providecommand \BibitemShut  [1]{\csname bibitem#1\endcsname}%
\let\auto@bib@innerbib\@empty
\bibitem [{\citenamefont {Schirotzek}\ \emph {et~al.}(2009)\citenamefont
  {Schirotzek}, \citenamefont {Wu}, \citenamefont {Sommer},\ and\ \citenamefont
  {Zwierlein}}]{Schirotzek2009}%
  \BibitemOpen
  \bibfield  {author} {\bibinfo {author} {\bibfnamefont {A.}~\bibnamefont
  {Schirotzek}}, \bibinfo {author} {\bibfnamefont {C.-H.}\ \bibnamefont {Wu}},
  \bibinfo {author} {\bibfnamefont {A.}~\bibnamefont {Sommer}}, \ and\ \bibinfo
  {author} {\bibfnamefont {M.~W.}\ \bibnamefont {Zwierlein}},\ }\href {\doibase
  10.1103/PhysRevLett.102.230402} {\bibfield  {journal} {\bibinfo  {journal}
  {Phys. Rev. Lett.}\ }\textbf {\bibinfo {volume} {102}},\ \bibinfo {pages}
  {230402} (\bibinfo {year} {2009})}\BibitemShut {NoStop}%
\bibitem [{\citenamefont {Greiner}\ \emph {et~al.}(2002)\citenamefont
  {Greiner}, \citenamefont {Mandel}, \citenamefont {Esslinger}, \citenamefont
  {H\"{a}nsch},\ and\ \citenamefont {Bloch}}]{Greiner2002}%
  \BibitemOpen
  \bibfield  {author} {\bibinfo {author} {\bibfnamefont {M.}~\bibnamefont
  {Greiner}}, \bibinfo {author} {\bibfnamefont {O.}~\bibnamefont {Mandel}},
  \bibinfo {author} {\bibfnamefont {T.}~\bibnamefont {Esslinger}}, \bibinfo
  {author} {\bibfnamefont {T.~W.}\ \bibnamefont {H\"{a}nsch}}, \ and\ \bibinfo
  {author} {\bibfnamefont {I.}~\bibnamefont {Bloch}},\ }\href {\doibase
  10.1038/415039a} {\bibfield  {journal} {\bibinfo  {journal} {Nature
  (London)}\ }\textbf {\bibinfo {volume} {415}},\ \bibinfo {pages} {39}
  (\bibinfo {year} {2002})}\BibitemShut {NoStop}%
\bibitem [{\citenamefont {Leggett}(1980)}]{Leggett1980}%
  \BibitemOpen
  \bibfield  {author} {\bibinfo {author} {\bibfnamefont {A.~J.}\ \bibnamefont
  {Leggett}},\ }in\ \href@noop {} {\emph {\bibinfo {booktitle} {Modern Trends
  in the Theory of Condensed Matter}}},\ \bibinfo {editor} {edited by\ \bibinfo
  {editor} {\bibfnamefont {A.}~\bibnamefont {Pekalski}}\ and\ \bibinfo {editor}
  {\bibfnamefont {J.}~\bibnamefont {Przystawa}}}\ (\bibinfo  {publisher}
  {Springer},\ \bibinfo {address} {Berlin},\ \bibinfo {year} {1980})\ pp.\
  \bibinfo {pages} {13--27}\BibitemShut {NoStop}%
\bibitem [{\citenamefont {Bourdel}\ \emph {et~al.}(2004)\citenamefont
  {Bourdel}, \citenamefont {Khaykovich}, \citenamefont {Cubizolles},
  \citenamefont {Zhang}, \citenamefont {Chevy}, \citenamefont {Teichmann},
  \citenamefont {Tarruell}, \citenamefont {Kokkelmans},\ and\ \citenamefont
  {Salomon}}]{Bourdel2004}%
  \BibitemOpen
  \bibfield  {author} {\bibinfo {author} {\bibfnamefont {T.}~\bibnamefont
  {Bourdel}}, \bibinfo {author} {\bibfnamefont {L.}~\bibnamefont {Khaykovich}},
  \bibinfo {author} {\bibfnamefont {J.}~\bibnamefont {Cubizolles}}, \bibinfo
  {author} {\bibfnamefont {J.}~\bibnamefont {Zhang}}, \bibinfo {author}
  {\bibfnamefont {F.}~\bibnamefont {Chevy}}, \bibinfo {author} {\bibfnamefont
  {M.}~\bibnamefont {Teichmann}}, \bibinfo {author} {\bibfnamefont
  {L.}~\bibnamefont {Tarruell}}, \bibinfo {author} {\bibfnamefont {S.~J. J.
  M.~F.}\ \bibnamefont {Kokkelmans}}, \ and\ \bibinfo {author} {\bibfnamefont
  {C.}~\bibnamefont {Salomon}},\ }\href {\doibase
  10.1103/PhysRevLett.93.050401} {\bibfield  {journal} {\bibinfo  {journal}
  {Phys. Rev. Lett.}\ }\textbf {\bibinfo {volume} {93}},\ \bibinfo {pages}
  {050401} (\bibinfo {year} {2004})}\BibitemShut {NoStop}%
\bibitem [{\citenamefont {Giorgini}\ \emph {et~al.}(2008)\citenamefont
  {Giorgini}, \citenamefont {Pitaevskii},\ and\ \citenamefont
  {Stringari}}]{Giorgini2008}%
  \BibitemOpen
  \bibfield  {author} {\bibinfo {author} {\bibfnamefont {S.}~\bibnamefont
  {Giorgini}}, \bibinfo {author} {\bibfnamefont {L.~P.}\ \bibnamefont
  {Pitaevskii}}, \ and\ \bibinfo {author} {\bibfnamefont {S.}~\bibnamefont
  {Stringari}},\ }\href {\doibase 10.1103/RevModPhys.80.1215} {\bibfield
  {journal} {\bibinfo  {journal} {Rev. Mod. Phys.}\ }\textbf {\bibinfo {volume}
  {80}},\ \bibinfo {pages} {1215} (\bibinfo {year} {2008})}\BibitemShut
  {NoStop}%
\bibitem [{\citenamefont {Gaunt}\ \emph {et~al.}(2013)\citenamefont {Gaunt},
  \citenamefont {Schmidutz}, \citenamefont {Gotlibovych}, \citenamefont
  {Smith},\ and\ \citenamefont {Hadzibabic}}]{Gaunt2013}%
  \BibitemOpen
  \bibfield  {author} {\bibinfo {author} {\bibfnamefont {A.~L.}\ \bibnamefont
  {Gaunt}}, \bibinfo {author} {\bibfnamefont {T.~F.}\ \bibnamefont
  {Schmidutz}}, \bibinfo {author} {\bibfnamefont {I.}~\bibnamefont
  {Gotlibovych}}, \bibinfo {author} {\bibfnamefont {R.~P.}\ \bibnamefont
  {Smith}}, \ and\ \bibinfo {author} {\bibfnamefont {Z.}~\bibnamefont
  {Hadzibabic}},\ }\href {\doibase 10.1103/PhysRevLett.110.200406} {\bibfield
  {journal} {\bibinfo  {journal} {Phys. Rev. Lett.}\ }\textbf {\bibinfo
  {volume} {110}},\ \bibinfo {pages} {200406} (\bibinfo {year}
  {2013})}\BibitemShut {NoStop}%
\bibitem [{\citenamefont {Chin}\ \emph {et~al.}(2010)\citenamefont {Chin},
  \citenamefont {Grimm}, \citenamefont {Julienne},\ and\ \citenamefont
  {Tiesinga}}]{Chin2010}%
  \BibitemOpen
  \bibfield  {author} {\bibinfo {author} {\bibfnamefont {C.}~\bibnamefont
  {Chin}}, \bibinfo {author} {\bibfnamefont {R.}~\bibnamefont {Grimm}},
  \bibinfo {author} {\bibfnamefont {P.}~\bibnamefont {Julienne}}, \ and\
  \bibinfo {author} {\bibfnamefont {E.}~\bibnamefont {Tiesinga}},\ }\href
  {\doibase 10.1103/RevModPhys.82.1225} {\bibfield  {journal} {\bibinfo
  {journal} {Rev. Mod. Phys.}\ }\textbf {\bibinfo {volume} {82}},\ \bibinfo
  {pages} {1225} (\bibinfo {year} {2010})}\BibitemShut {NoStop}%
\bibitem [{\citenamefont {Baker}(1999)}]{Baker1999}%
  \BibitemOpen
  \bibfield  {author} {\bibinfo {author} {\bibfnamefont {G.~A.}\ \bibnamefont
  {Baker}},\ }\href {\doibase 10.1103/PhysRevC.60.054311} {\bibfield  {journal}
  {\bibinfo  {journal} {Phys. Rev. C}\ }\textbf {\bibinfo {volume} {60}},\
  \bibinfo {pages} {054311} (\bibinfo {year} {1999})}\BibitemShut {NoStop}%
\bibitem [{\citenamefont {Landau}\ and\ \citenamefont
  {Lifshitz}(1981)}]{Landau1981}%
  \BibitemOpen
  \bibfield  {author} {\bibinfo {author} {\bibfnamefont {L.~D.}\ \bibnamefont
  {Landau}}\ and\ \bibinfo {author} {\bibfnamefont {E.~M.}\ \bibnamefont
  {Lifshitz}},\ }\href@noop {} {\emph {\bibinfo {title} {{Quantum Mechanics:
  Non-Relativistic Theory}}}}\ (\bibinfo  {publisher} {Butterworth-Heinemann},\
  \bibinfo {address} {Oxford, UK},\ \bibinfo {year} {1981})\BibitemShut
  {NoStop}%
\bibitem [{\citenamefont {Miller}\ \emph {et~al.}(1990)\citenamefont {Miller},
  \citenamefont {Nefkens},\ and\ \citenamefont {\v{S}laus}}]{Miller1990}%
  \BibitemOpen
  \bibfield  {author} {\bibinfo {author} {\bibfnamefont {G.}~\bibnamefont
  {Miller}}, \bibinfo {author} {\bibfnamefont {B.}~\bibnamefont {Nefkens}}, \
  and\ \bibinfo {author} {\bibfnamefont {I.}~\bibnamefont {\v{S}laus}},\ }\href
  {\doibase 10.1016/0370-1573(90)90102-8} {\bibfield  {journal} {\bibinfo
  {journal} {Phys. Rep.}\ }\textbf {\bibinfo {volume} {194}},\ \bibinfo {pages}
  {1} (\bibinfo {year} {1990})}\BibitemShut {NoStop}%
\bibitem [{\citenamefont {Forbes}\ \emph {et~al.}(2012)\citenamefont {Forbes},
  \citenamefont {Gandolfi},\ and\ \citenamefont {Gezerlis}}]{Forbes2012}%
  \BibitemOpen
  \bibfield  {author} {\bibinfo {author} {\bibfnamefont {M.~M.}\ \bibnamefont
  {Forbes}}, \bibinfo {author} {\bibfnamefont {S.}~\bibnamefont {Gandolfi}}, \
  and\ \bibinfo {author} {\bibfnamefont {A.}~\bibnamefont {Gezerlis}},\ }\href
  {\doibase 10.1103/PhysRevA.86.053603} {\bibfield  {journal} {\bibinfo
  {journal} {Phys. Rev. A}\ }\textbf {\bibinfo {volume} {86}},\ \bibinfo
  {pages} {053603} (\bibinfo {year} {2012})}\BibitemShut {NoStop}%
\bibitem [{\citenamefont {Gurarie}\ and\ \citenamefont
  {Radzihovsky}(2007)}]{Gurarie2007}%
  \BibitemOpen
  \bibfield  {author} {\bibinfo {author} {\bibfnamefont {V.}~\bibnamefont
  {Gurarie}}\ and\ \bibinfo {author} {\bibfnamefont {L.}~\bibnamefont
  {Radzihovsky}},\ }\href {\doibase 10.1016/j.aop.2006.10.009} {\bibfield
  {journal} {\bibinfo  {journal} {Ann. Phys. (NY)}\ }\textbf {\bibinfo {volume}
  {322}},\ \bibinfo {pages} {2} (\bibinfo {year} {2007})}\BibitemShut {NoStop}%
\bibitem [{\citenamefont {Wang}\ \emph {et~al.}(2011)\citenamefont {Wang},
  \citenamefont {D'Incao},\ and\ \citenamefont {Esry}}]{Wang2011}%
  \BibitemOpen
  \bibfield  {author} {\bibinfo {author} {\bibfnamefont {Y.}~\bibnamefont
  {Wang}}, \bibinfo {author} {\bibfnamefont {J.~P.}\ \bibnamefont {D'Incao}}, \
  and\ \bibinfo {author} {\bibfnamefont {B.~D.}\ \bibnamefont {Esry}},\ }\href
  {\doibase 10.1103/PhysRevA.83.042710} {\bibfield  {journal} {\bibinfo
  {journal} {Phys. Rev. A}\ }\textbf {\bibinfo {volume} {83}},\ \bibinfo
  {pages} {042710} (\bibinfo {year} {2011})}\BibitemShut {NoStop}%
\bibitem [{\citenamefont {Li}\ \emph {et~al.}(2011)\citenamefont {Li},
  \citenamefont {Koloren\v{c}},\ and\ \citenamefont {Mitas}}]{Li2011}%
  \BibitemOpen
  \bibfield  {author} {\bibinfo {author} {\bibfnamefont {X.}~\bibnamefont
  {Li}}, \bibinfo {author} {\bibfnamefont {J.}~\bibnamefont {Koloren\v{c}}}, \
  and\ \bibinfo {author} {\bibfnamefont {L.}~\bibnamefont {Mitas}},\ }\href
  {\doibase 10.1103/PhysRevA.84.023615} {\bibfield  {journal} {\bibinfo
  {journal} {Phys. Rev. A}\ }\textbf {\bibinfo {volume} {84}},\ \bibinfo
  {pages} {023615} (\bibinfo {year} {2011})}\BibitemShut {NoStop}%
\bibitem [{\citenamefont {Gandolfi}\ \emph {et~al.}(2011)\citenamefont
  {Gandolfi}, \citenamefont {Schmidt},\ and\ \citenamefont
  {Carlson}}]{Gandolfi2011}%
  \BibitemOpen
  \bibfield  {author} {\bibinfo {author} {\bibfnamefont {S.}~\bibnamefont
  {Gandolfi}}, \bibinfo {author} {\bibfnamefont {K.~E.}\ \bibnamefont
  {Schmidt}}, \ and\ \bibinfo {author} {\bibfnamefont {J.}~\bibnamefont
  {Carlson}},\ }\href {\doibase 10.1103/PhysRevA.83.041601} {\bibfield
  {journal} {\bibinfo  {journal} {Phys. Rev. A}\ }\textbf {\bibinfo {volume}
  {83}},\ \bibinfo {pages} {041601} (\bibinfo {year} {2011})}\BibitemShut
  {NoStop}%
\bibitem [{\citenamefont {Forbes}\ \emph {et~al.}(2011)\citenamefont {Forbes},
  \citenamefont {Gandolfi},\ and\ \citenamefont {Gezerlis}}]{Forbes2011}%
  \BibitemOpen
  \bibfield  {author} {\bibinfo {author} {\bibfnamefont {M.~M.}\ \bibnamefont
  {Forbes}}, \bibinfo {author} {\bibfnamefont {S.}~\bibnamefont {Gandolfi}}, \
  and\ \bibinfo {author} {\bibfnamefont {A.}~\bibnamefont {Gezerlis}},\ }\href
  {\doibase 10.1103/PhysRevLett.106.235303} {\bibfield  {journal} {\bibinfo
  {journal} {Phys. Rev. Lett.}\ }\textbf {\bibinfo {volume} {106}},\ \bibinfo
  {pages} {235303} (\bibinfo {year} {2011})}\BibitemShut {NoStop}%
\bibitem [{\citenamefont {{De Palo}}\ \emph {et~al.}(2004)\citenamefont {{De
  Palo}}, \citenamefont {Chiofalo}, \citenamefont {Holland},\ and\
  \citenamefont {Kokkelmans}}]{DePalo2004}%
  \BibitemOpen
  \bibfield  {author} {\bibinfo {author} {\bibfnamefont {S.}~\bibnamefont {{De
  Palo}}}, \bibinfo {author} {\bibfnamefont {M.}~\bibnamefont {Chiofalo}},
  \bibinfo {author} {\bibfnamefont {M.}~\bibnamefont {Holland}}, \ and\
  \bibinfo {author} {\bibfnamefont {S.}~\bibnamefont {Kokkelmans}},\ }\href
  {\doibase 10.1016/j.physleta.2004.05.034} {\bibfield  {journal} {\bibinfo
  {journal} {Phys. Lett. A}\ }\textbf {\bibinfo {volume} {327}},\ \bibinfo
  {pages} {490} (\bibinfo {year} {2004})}\BibitemShut {NoStop}%
\bibitem [{\citenamefont {Jensen}\ \emph {et~al.}(2006)\citenamefont {Jensen},
  \citenamefont {Nilsen},\ and\ \citenamefont {Watanabe}}]{Jensen2006}%
  \BibitemOpen
  \bibfield  {author} {\bibinfo {author} {\bibfnamefont {L.~M.}\ \bibnamefont
  {Jensen}}, \bibinfo {author} {\bibfnamefont {H.~M.}\ \bibnamefont {Nilsen}},
  \ and\ \bibinfo {author} {\bibfnamefont {G.}~\bibnamefont {Watanabe}},\
  }\href {\doibase 10.1103/PhysRevA.74.043608} {\bibfield  {journal} {\bibinfo
  {journal} {Phys. Rev. A}\ }\textbf {\bibinfo {volume} {74}},\ \bibinfo
  {pages} {043608} (\bibinfo {year} {2006})}\BibitemShut {NoStop}%
\bibitem [{\citenamefont {{De Palo}}\ \emph {et~al.}(2005)\citenamefont {{De
  Palo}}, \citenamefont {Chiofalo}, \citenamefont {Holland},\ and\
  \citenamefont {Kokkelmans}}]{DePalo2005}%
  \BibitemOpen
  \bibfield  {author} {\bibinfo {author} {\bibfnamefont {S.}~\bibnamefont {{De
  Palo}}}, \bibinfo {author} {\bibfnamefont {M.~L.}\ \bibnamefont {Chiofalo}},
  \bibinfo {author} {\bibfnamefont {M.~J.}\ \bibnamefont {Holland}}, \ and\
  \bibinfo {author} {\bibfnamefont {S.}~\bibnamefont {Kokkelmans}},\
  }\href@noop {} {\bibfield  {journal} {\bibinfo  {journal} {Laser Phys.}\
  }\textbf {\bibinfo {volume} {15}},\ \bibinfo {pages} {376} (\bibinfo {year}
  {2005})}\BibitemShut {NoStop}%
\bibitem [{\citenamefont {Timmermans}\ \emph {et~al.}(1999)\citenamefont
  {Timmermans}, \citenamefont {Tomassini}, \citenamefont {Hussein},\ and\
  \citenamefont {Kerman}}]{Timmermans1999}%
  \BibitemOpen
  \bibfield  {author} {\bibinfo {author} {\bibfnamefont {E.}~\bibnamefont
  {Timmermans}}, \bibinfo {author} {\bibfnamefont {T.}~\bibnamefont
  {Tomassini}}, \bibinfo {author} {\bibfnamefont {M.}~\bibnamefont {Hussein}},
  \ and\ \bibinfo {author} {\bibfnamefont {A.}~\bibnamefont {Kerman}},\ }\href
  {\doibase 10.1016/S0370-1573(99)00025-3} {\bibfield  {journal} {\bibinfo
  {journal} {Phys. Rep.}\ }\textbf {\bibinfo {volume} {315}},\ \bibinfo {pages}
  {199} (\bibinfo {year} {1999})}\BibitemShut {NoStop}%
\bibitem [{\citenamefont {Bugnion}\ \emph {et~al.}(2014)\citenamefont
  {Bugnion}, \citenamefont {{L\'{o}pez R\'{\i}os}}, \citenamefont {Needs},\
  and\ \citenamefont {Conduit}}]{Bugnion2014}%
  \BibitemOpen
  \bibfield  {author} {\bibinfo {author} {\bibfnamefont {P.~O.}\ \bibnamefont
  {Bugnion}}, \bibinfo {author} {\bibfnamefont {P.}~\bibnamefont {{L\'{o}pez
  R\'{\i}os}}}, \bibinfo {author} {\bibfnamefont {R.~J.}\ \bibnamefont
  {Needs}}, \ and\ \bibinfo {author} {\bibfnamefont {G.~J.}\ \bibnamefont
  {Conduit}},\ }\href {\doibase 10.1103/PhysRevA.90.033626} {\bibfield
  {journal} {\bibinfo  {journal} {Phys. Rev. A}\ }\textbf {\bibinfo {volume}
  {90}},\ \bibinfo {pages} {033626} (\bibinfo {year} {2014})}\BibitemShut
  {NoStop}%
\bibitem [{\citenamefont {Whitehead}\ \emph
  {et~al.}(2016{\natexlab{a}})\citenamefont {Whitehead}, \citenamefont
  {Schonenberg}, \citenamefont {Kongsuwan}, \citenamefont {Needs},\ and\
  \citenamefont {Conduit}}]{Whitehead2016a}%
  \BibitemOpen
  \bibfield  {author} {\bibinfo {author} {\bibfnamefont {T.~M.}\ \bibnamefont
  {Whitehead}}, \bibinfo {author} {\bibfnamefont {L.~M.}\ \bibnamefont
  {Schonenberg}}, \bibinfo {author} {\bibfnamefont {N.}~\bibnamefont
  {Kongsuwan}}, \bibinfo {author} {\bibfnamefont {R.~J.}\ \bibnamefont
  {Needs}}, \ and\ \bibinfo {author} {\bibfnamefont {G.~J.}\ \bibnamefont
  {Conduit}},\ }\href {\doibase 10.1103/PhysRevA.93.042702} {\bibfield
  {journal} {\bibinfo  {journal} {Phys. Rev. A}\ }\textbf {\bibinfo {volume}
  {93}},\ \bibinfo {pages} {042702} (\bibinfo {year}
  {2016}{\natexlab{a}})}\BibitemShut {NoStop}%
\bibitem [{\citenamefont {Foulkes}\ \emph {et~al.}(2001)\citenamefont
  {Foulkes}, \citenamefont {Mitas}, \citenamefont {Needs},\ and\ \citenamefont
  {Rajagopal}}]{Foulkes2001}%
  \BibitemOpen
  \bibfield  {author} {\bibinfo {author} {\bibfnamefont {W.~M.~C.}\
  \bibnamefont {Foulkes}}, \bibinfo {author} {\bibfnamefont {L.}~\bibnamefont
  {Mitas}}, \bibinfo {author} {\bibfnamefont {R.~J.}\ \bibnamefont {Needs}}, \
  and\ \bibinfo {author} {\bibfnamefont {G.}~\bibnamefont {Rajagopal}},\ }\href
  {\doibase 10.1103/RevModPhys.73.33} {\bibfield  {journal} {\bibinfo
  {journal} {Rev. Mod. Phys.}\ }\textbf {\bibinfo {volume} {73}},\ \bibinfo
  {pages} {33} (\bibinfo {year} {2001})}\BibitemShut {NoStop}%
\bibitem [{\citenamefont {Chadan}\ and\ \citenamefont
  {Sabatier}(1989)}]{Chadan1989}%
  \BibitemOpen
  \bibfield  {author} {\bibinfo {author} {\bibfnamefont {K.}~\bibnamefont
  {Chadan}}\ and\ \bibinfo {author} {\bibfnamefont {P.~C.}\ \bibnamefont
  {Sabatier}},\ }\href@noop {} {\emph {\bibinfo {title} {{Inverse Problems in
  Quantum Scattering Theory}}}}\ (\bibinfo  {publisher} {Springer},\ \bibinfo
  {address} {New York},\ \bibinfo {year} {1989})\BibitemShut {NoStop}%
\bibitem [{\citenamefont {Lloyd-Williams}\ \emph {et~al.}(2015)\citenamefont
  {Lloyd-Williams}, \citenamefont {Needs},\ and\ \citenamefont
  {Conduit}}]{Lloyd-Williams2015}%
  \BibitemOpen
  \bibfield  {author} {\bibinfo {author} {\bibfnamefont {J.~H.}\ \bibnamefont
  {Lloyd-Williams}}, \bibinfo {author} {\bibfnamefont {R.~J.}\ \bibnamefont
  {Needs}}, \ and\ \bibinfo {author} {\bibfnamefont {G.~J.}\ \bibnamefont
  {Conduit}},\ }\href {\doibase 10.1103/PhysRevB.92.075106} {\bibfield
  {journal} {\bibinfo  {journal} {Phys. Rev. B}\ }\textbf {\bibinfo {volume}
  {92}},\ \bibinfo {pages} {075106} (\bibinfo {year} {2015})}\BibitemShut
  {NoStop}%
\bibitem [{\citenamefont {Whitehead}\ and\ \citenamefont
  {Conduit}(2016)}]{Whitehead2016}%
  \BibitemOpen
  \bibfield  {author} {\bibinfo {author} {\bibfnamefont {T.~M.}\ \bibnamefont
  {Whitehead}}\ and\ \bibinfo {author} {\bibfnamefont {G.~J.}\ \bibnamefont
  {Conduit}},\ }\href {\doibase 10.1103/PhysRevA.93.022706} {\bibfield
  {journal} {\bibinfo  {journal} {Phys. Rev. A}\ }\textbf {\bibinfo {volume}
  {93}},\ \bibinfo {pages} {022706} (\bibinfo {year} {2016})}\BibitemShut
  {NoStop}%
\bibitem [{\citenamefont {Schonenberg}\ and\ \citenamefont
  {Conduit}(2017)}]{Schonenberg2016a}%
  \BibitemOpen
  \bibfield  {author} {\bibinfo {author} {\bibfnamefont {L.~M.}\ \bibnamefont
  {Schonenberg}}\ and\ \bibinfo {author} {\bibfnamefont {G.~J.}\ \bibnamefont
  {Conduit}},\ }\href {https://doi.org/10.17863/CAM.7090} {\bibfield  {journal}
  {\bibinfo  {journal} {Cambridge Univ. Apollo Repos.
  https://doi.org/10.17863/CAM.7090}\ } (\bibinfo {year} {2017})}\BibitemShut
  {NoStop}%
\bibitem [{\citenamefont {Astrakharchik}\ \emph {et~al.}(2004)\citenamefont
  {Astrakharchik}, \citenamefont {Boronat}, \citenamefont {Casulleras},\ and\
  \citenamefont {Giorgini}}]{Astrakharchik2004}%
  \BibitemOpen
  \bibfield  {author} {\bibinfo {author} {\bibfnamefont {G.~E.}\ \bibnamefont
  {Astrakharchik}}, \bibinfo {author} {\bibfnamefont {J.}~\bibnamefont
  {Boronat}}, \bibinfo {author} {\bibfnamefont {J.}~\bibnamefont {Casulleras}},
  \ and\ \bibinfo {author} {\bibfnamefont {S.}~\bibnamefont {Giorgini}},\
  }\href {\doibase 10.1103/PhysRevLett.93.200404} {\bibfield  {journal}
  {\bibinfo  {journal} {Phys. Rev. Lett.}\ }\textbf {\bibinfo {volume} {93}},\
  \bibinfo {pages} {200404} (\bibinfo {year} {2004})}\BibitemShut {NoStop}%
\bibitem [{\citenamefont {Astrakharchik}\ \emph {et~al.}(2005)\citenamefont
  {Astrakharchik}, \citenamefont {Boronat}, \citenamefont {Casulleras},\ and\
  \citenamefont {Giorgini}}]{Astrakharchik2005}%
  \BibitemOpen
  \bibfield  {author} {\bibinfo {author} {\bibfnamefont {G.~E.}\ \bibnamefont
  {Astrakharchik}}, \bibinfo {author} {\bibfnamefont {J.}~\bibnamefont
  {Boronat}}, \bibinfo {author} {\bibfnamefont {J.}~\bibnamefont {Casulleras}},
  \ and\ \bibinfo {author} {\bibfnamefont {S.}~\bibnamefont {Giorgini}},\
  }\href {\doibase 10.1103/PhysRevLett.95.230405} {\bibfield  {journal}
  {\bibinfo  {journal} {Phys. Rev. Lett.}\ }\textbf {\bibinfo {volume} {95}},\
  \bibinfo {pages} {230405} (\bibinfo {year} {2005})}\BibitemShut {NoStop}%
\bibitem [{\citenamefont {Morris}\ \emph {et~al.}(2010)\citenamefont {Morris},
  \citenamefont {{L\'{o}pez R\'{\i}os}},\ and\ \citenamefont
  {Needs}}]{Morris2010}%
  \BibitemOpen
  \bibfield  {author} {\bibinfo {author} {\bibfnamefont {A.~J.}\ \bibnamefont
  {Morris}}, \bibinfo {author} {\bibfnamefont {P.}~\bibnamefont {{L\'{o}pez
  R\'{\i}os}}}, \ and\ \bibinfo {author} {\bibfnamefont {R.~J.}\ \bibnamefont
  {Needs}},\ }\href {\doibase 10.1103/PhysRevA.81.033619} {\bibfield  {journal}
  {\bibinfo  {journal} {Phys. Rev. A}\ }\textbf {\bibinfo {volume} {81}},\
  \bibinfo {pages} {033619} (\bibinfo {year} {2010})}\BibitemShut {NoStop}%
\bibitem [{\citenamefont {Carlson}\ \emph {et~al.}(2011)\citenamefont
  {Carlson}, \citenamefont {Gandolfi}, \citenamefont {Schmidt},\ and\
  \citenamefont {Zhang}}]{Carlson2011}%
  \BibitemOpen
  \bibfield  {author} {\bibinfo {author} {\bibfnamefont {J.}~\bibnamefont
  {Carlson}}, \bibinfo {author} {\bibfnamefont {S.}~\bibnamefont {Gandolfi}},
  \bibinfo {author} {\bibfnamefont {K.~E.}\ \bibnamefont {Schmidt}}, \ and\
  \bibinfo {author} {\bibfnamefont {S.}~\bibnamefont {Zhang}},\ }\href
  {\doibase 10.1103/PhysRevA.84.061602} {\bibfield  {journal} {\bibinfo
  {journal} {Phys. Rev. A}\ }\textbf {\bibinfo {volume} {84}},\ \bibinfo
  {pages} {061602} (\bibinfo {year} {2011})}\BibitemShut {NoStop}%
\bibitem [{\citenamefont {Ceperley}\ and\ \citenamefont
  {Alder}(1980)}]{Ceperley1980}%
  \BibitemOpen
  \bibfield  {author} {\bibinfo {author} {\bibfnamefont {D.~M.}\ \bibnamefont
  {Ceperley}}\ and\ \bibinfo {author} {\bibfnamefont {B.~J.}\ \bibnamefont
  {Alder}},\ }\href {\doibase 10.1103/PhysRevLett.45.566} {\bibfield  {journal}
  {\bibinfo  {journal} {Phys. Rev. Lett.}\ }\textbf {\bibinfo {volume} {45}},\
  \bibinfo {pages} {566} (\bibinfo {year} {1980})}\BibitemShut {NoStop}%
\bibitem [{\citenamefont {Umrigar}\ \emph {et~al.}(1993)\citenamefont
  {Umrigar}, \citenamefont {Nightingale},\ and\ \citenamefont
  {Runge}}]{Umrigar1993}%
  \BibitemOpen
  \bibfield  {author} {\bibinfo {author} {\bibfnamefont {C.~J.}\ \bibnamefont
  {Umrigar}}, \bibinfo {author} {\bibfnamefont {M.~P.}\ \bibnamefont
  {Nightingale}}, \ and\ \bibinfo {author} {\bibfnamefont {K.~J.}\ \bibnamefont
  {Runge}},\ }\href {\doibase 10.1063/1.465195} {\bibfield  {journal} {\bibinfo
   {journal} {J. Chem. Phys.}\ }\textbf {\bibinfo {volume} {99}},\ \bibinfo
  {pages} {2865} (\bibinfo {year} {1993})}\BibitemShut {NoStop}%
\bibitem [{\citenamefont {Needs}\ \emph {et~al.}(2010)\citenamefont {Needs},
  \citenamefont {Towler}, \citenamefont {Drummond},\ and\ \citenamefont
  {{L\'{o}pez R\'{\i}os}}}]{Needs2010}%
  \BibitemOpen
  \bibfield  {author} {\bibinfo {author} {\bibfnamefont {R.~J.}\ \bibnamefont
  {Needs}}, \bibinfo {author} {\bibfnamefont {M.~D.}\ \bibnamefont {Towler}},
  \bibinfo {author} {\bibfnamefont {N.~D.}\ \bibnamefont {Drummond}}, \ and\
  \bibinfo {author} {\bibfnamefont {P.}~\bibnamefont {{L\'{o}pez R\'{\i}os}}},\
  }\href {\doibase 10.1088/0953-8984/22/2/023201} {\bibfield  {journal}
  {\bibinfo  {journal} {J. Phys. Condens. Matter}\ }\textbf {\bibinfo {volume}
  {22}},\ \bibinfo {pages} {023201} (\bibinfo {year} {2010})}\BibitemShut
  {NoStop}%
\bibitem [{\citenamefont {Carlson}\ \emph {et~al.}(2003)\citenamefont
  {Carlson}, \citenamefont {Chang}, \citenamefont {Pandharipande},\ and\
  \citenamefont {Schmidt}}]{Carlson2003}%
  \BibitemOpen
  \bibfield  {author} {\bibinfo {author} {\bibfnamefont {J.}~\bibnamefont
  {Carlson}}, \bibinfo {author} {\bibfnamefont {S.-Y.}\ \bibnamefont {Chang}},
  \bibinfo {author} {\bibfnamefont {V.~R.}\ \bibnamefont {Pandharipande}}, \
  and\ \bibinfo {author} {\bibfnamefont {K.~E.}\ \bibnamefont {Schmidt}},\
  }\href {\doibase 10.1103/PhysRevLett.91.050401} {\bibfield  {journal}
  {\bibinfo  {journal} {Phys. Rev. Lett.}\ }\textbf {\bibinfo {volume} {91}},\
  \bibinfo {pages} {050401} (\bibinfo {year} {2003})}\BibitemShut {NoStop}%
\bibitem [{\citenamefont {{L\'{o}pez R\'{\i}os}}\ \emph
  {et~al.}(2006)\citenamefont {{L\'{o}pez R\'{\i}os}}, \citenamefont {Ma},
  \citenamefont {Drummond}, \citenamefont {Towler},\ and\ \citenamefont
  {Needs}}]{LopezRios2006}%
  \BibitemOpen
  \bibfield  {author} {\bibinfo {author} {\bibfnamefont {P.}~\bibnamefont
  {{L\'{o}pez R\'{\i}os}}}, \bibinfo {author} {\bibfnamefont {A.}~\bibnamefont
  {Ma}}, \bibinfo {author} {\bibfnamefont {N.~D.}\ \bibnamefont {Drummond}},
  \bibinfo {author} {\bibfnamefont {M.~D.}\ \bibnamefont {Towler}}, \ and\
  \bibinfo {author} {\bibfnamefont {R.~J.}\ \bibnamefont {Needs}},\ }\href
  {\doibase 10.1103/PhysRevE.74.066701} {\bibfield  {journal} {\bibinfo
  {journal} {Phys. Rev. E}\ }\textbf {\bibinfo {volume} {74}},\ \bibinfo
  {pages} {066701} (\bibinfo {year} {2006})}\BibitemShut {NoStop}%
\bibitem [{\citenamefont {Drummond}\ \emph {et~al.}(2004)\citenamefont
  {Drummond}, \citenamefont {Towler},\ and\ \citenamefont
  {Needs}}]{Drummond2004}%
  \BibitemOpen
  \bibfield  {author} {\bibinfo {author} {\bibfnamefont {N.~D.}\ \bibnamefont
  {Drummond}}, \bibinfo {author} {\bibfnamefont {M.~D.}\ \bibnamefont
  {Towler}}, \ and\ \bibinfo {author} {\bibfnamefont {R.~J.}\ \bibnamefont
  {Needs}},\ }\href {\doibase 10.1103/PhysRevB.70.235119} {\bibfield  {journal}
  {\bibinfo  {journal} {Phys. Rev. B}\ }\textbf {\bibinfo {volume} {70}},\
  \bibinfo {pages} {235119} (\bibinfo {year} {2004})}\BibitemShut {NoStop}%
\bibitem [{\citenamefont {Whitehead}\ \emph
  {et~al.}(2016{\natexlab{b}})\citenamefont {Whitehead}, \citenamefont
  {Michael},\ and\ \citenamefont {Conduit}}]{Whitehead2016c}%
  \BibitemOpen
  \bibfield  {author} {\bibinfo {author} {\bibfnamefont {T.~M.}\ \bibnamefont
  {Whitehead}}, \bibinfo {author} {\bibfnamefont {M.~H.}\ \bibnamefont
  {Michael}}, \ and\ \bibinfo {author} {\bibfnamefont {G.~J.}\ \bibnamefont
  {Conduit}},\ }\href {\doibase 10.1103/PhysRevB.94.035157} {\bibfield
  {journal} {\bibinfo  {journal} {Phys. Rev. B}\ }\textbf {\bibinfo {volume}
  {94}},\ \bibinfo {pages} {035157} (\bibinfo {year}
  {2016}{\natexlab{b}})}\BibitemShut {NoStop}%
\bibitem [{\citenamefont {Ceperley}\ and\ \citenamefont
  {Kalos}(1986)}]{Ceperley1986}%
  \BibitemOpen
  \bibfield  {author} {\bibinfo {author} {\bibfnamefont {D.~M.}\ \bibnamefont
  {Ceperley}}\ and\ \bibinfo {author} {\bibfnamefont {M.~H.}\ \bibnamefont
  {Kalos}},\ }in\ \href {\doibase 10.1007/978-3-642-82803-4\_4} {\emph
  {\bibinfo {booktitle} {Monte Carlo Methods in Statistical Physics}}},\
  \bibinfo {editor} {edited by\ \bibinfo {editor} {\bibfnamefont
  {K.}~\bibnamefont {Binder}}}\ (\bibinfo  {publisher} {Springer},\ \bibinfo
  {address} {Berlin; Heidelberg},\ \bibinfo {year} {1986})\ \bibinfo {edition}
  {2nd}\ ed.,\ Chap.~\bibinfo {chapter} {4}, pp.\ \bibinfo {pages}
  {145--194}\BibitemShut {NoStop}%
\bibitem [{\citenamefont {Mella}\ \emph {et~al.}(2000)\citenamefont {Mella},
  \citenamefont {Morosi},\ and\ \citenamefont {Bressanini}}]{Mella2000}%
  \BibitemOpen
  \bibfield  {author} {\bibinfo {author} {\bibfnamefont {M.}~\bibnamefont
  {Mella}}, \bibinfo {author} {\bibfnamefont {G.}~\bibnamefont {Morosi}}, \
  and\ \bibinfo {author} {\bibfnamefont {D.}~\bibnamefont {Bressanini}},\
  }\href {\doibase 10.1103/PhysRevE.61.2050} {\bibfield  {journal} {\bibinfo
  {journal} {Phys. Rev. E}\ }\textbf {\bibinfo {volume} {61}},\ \bibinfo
  {pages} {2050} (\bibinfo {year} {2000})}\BibitemShut {NoStop}%
\bibitem [{\citenamefont {Sarsa}\ \emph {et~al.}(2002)\citenamefont {Sarsa},
  \citenamefont {Boronat},\ and\ \citenamefont {Casulleras}}]{Sarsa2002}%
  \BibitemOpen
  \bibfield  {author} {\bibinfo {author} {\bibfnamefont {A.}~\bibnamefont
  {Sarsa}}, \bibinfo {author} {\bibfnamefont {J.}~\bibnamefont {Boronat}}, \
  and\ \bibinfo {author} {\bibfnamefont {J.}~\bibnamefont {Casulleras}},\
  }\href {\doibase 10.1063/1.1446847} {\bibfield  {journal} {\bibinfo
  {journal} {J. Chem. Phys.}\ }\textbf {\bibinfo {volume} {116}},\ \bibinfo
  {pages} {5956} (\bibinfo {year} {2002})}\BibitemShut {NoStop}%
\bibitem [{\citenamefont {Pessoa}\ \emph {et~al.}(2015)\citenamefont {Pessoa},
  \citenamefont {Gandolfi}, \citenamefont {Vitiello},\ and\ \citenamefont
  {Schmidt}}]{Pessoa2015}%
  \BibitemOpen
  \bibfield  {author} {\bibinfo {author} {\bibfnamefont {R.}~\bibnamefont
  {Pessoa}}, \bibinfo {author} {\bibfnamefont {S.}~\bibnamefont {Gandolfi}},
  \bibinfo {author} {\bibfnamefont {S.~A.}\ \bibnamefont {Vitiello}}, \ and\
  \bibinfo {author} {\bibfnamefont {K.~E.}\ \bibnamefont {Schmidt}},\ }\href
  {\doibase 10.1103/PhysRevA.92.063625} {\bibfield  {journal} {\bibinfo
  {journal} {Phys. Rev. A}\ }\textbf {\bibinfo {volume} {92}},\ \bibinfo
  {pages} {063625} (\bibinfo {year} {2015})}\BibitemShut {NoStop}%
\bibitem [{\citenamefont {Luo}\ and\ \citenamefont {Thomas}(2009)}]{Luo2009}%
  \BibitemOpen
  \bibfield  {author} {\bibinfo {author} {\bibfnamefont {L.}~\bibnamefont
  {Luo}}\ and\ \bibinfo {author} {\bibfnamefont {J.~E.}\ \bibnamefont
  {Thomas}},\ }\href {\doibase 10.1007/s10909-008-9850-2} {\bibfield  {journal}
  {\bibinfo  {journal} {J. Low Temp. Phys.}\ }\textbf {\bibinfo {volume}
  {154}},\ \bibinfo {pages} {1} (\bibinfo {year} {2009})}\BibitemShut {NoStop}%
\bibitem [{\citenamefont {Navon}\ \emph {et~al.}(2010)\citenamefont {Navon},
  \citenamefont {Nascimbene}, \citenamefont {Chevy},\ and\ \citenamefont
  {Salomon}}]{Navon2010}%
  \BibitemOpen
  \bibfield  {author} {\bibinfo {author} {\bibfnamefont {N.}~\bibnamefont
  {Navon}}, \bibinfo {author} {\bibfnamefont {S.}~\bibnamefont {Nascimbene}},
  \bibinfo {author} {\bibfnamefont {F.}~\bibnamefont {Chevy}}, \ and\ \bibinfo
  {author} {\bibfnamefont {C.}~\bibnamefont {Salomon}},\ }\href {\doibase
  10.1126/science.1187582} {\bibfield  {journal} {\bibinfo  {journal}
  {Science}\ }\textbf {\bibinfo {volume} {328}},\ \bibinfo {pages} {729}
  (\bibinfo {year} {2010})}\BibitemShut {NoStop}%
\bibitem [{\citenamefont {Z\"{u}rn}\ \emph {et~al.}(2013)\citenamefont
  {Z\"{u}rn}, \citenamefont {Lompe}, \citenamefont {Wenz}, \citenamefont
  {Jochim}, \citenamefont {Julienne},\ and\ \citenamefont {Hutson}}]{Zurn2013}%
  \BibitemOpen
  \bibfield  {author} {\bibinfo {author} {\bibfnamefont {G.}~\bibnamefont
  {Z\"{u}rn}}, \bibinfo {author} {\bibfnamefont {T.}~\bibnamefont {Lompe}},
  \bibinfo {author} {\bibfnamefont {A.~N.}\ \bibnamefont {Wenz}}, \bibinfo
  {author} {\bibfnamefont {S.}~\bibnamefont {Jochim}}, \bibinfo {author}
  {\bibfnamefont {P.~S.}\ \bibnamefont {Julienne}}, \ and\ \bibinfo {author}
  {\bibfnamefont {J.~M.}\ \bibnamefont {Hutson}},\ }\href {\doibase
  10.1103/PhysRevLett.110.135301} {\bibfield  {journal} {\bibinfo  {journal}
  {Phys. Rev. Lett.}\ }\textbf {\bibinfo {volume} {110}},\ \bibinfo {pages}
  {135301} (\bibinfo {year} {2013})}\BibitemShut {NoStop}%
\bibitem [{\citenamefont {Ku}\ \emph {et~al.}(2012)\citenamefont {Ku},
  \citenamefont {Sommer}, \citenamefont {Cheuk},\ and\ \citenamefont
  {Zwierlein}}]{Ku2012}%
  \BibitemOpen
  \bibfield  {author} {\bibinfo {author} {\bibfnamefont {M.~J.~H.}\
  \bibnamefont {Ku}}, \bibinfo {author} {\bibfnamefont {A.~T.}\ \bibnamefont
  {Sommer}}, \bibinfo {author} {\bibfnamefont {L.~W.}\ \bibnamefont {Cheuk}}, \
  and\ \bibinfo {author} {\bibfnamefont {M.~W.}\ \bibnamefont {Zwierlein}},\
  }\href {\doibase 10.1126/science.1214987} {\bibfield  {journal} {\bibinfo
  {journal} {Science}\ }\textbf {\bibinfo {volume} {335}},\ \bibinfo {pages}
  {563} (\bibinfo {year} {2012})}\BibitemShut {NoStop}%
\bibitem [{\citenamefont {Werner}\ and\ \citenamefont
  {Castin}(2012)}]{Werner2012}%
  \BibitemOpen
  \bibfield  {author} {\bibinfo {author} {\bibfnamefont {F.}~\bibnamefont
  {Werner}}\ and\ \bibinfo {author} {\bibfnamefont {Y.}~\bibnamefont
  {Castin}},\ }\href {\doibase 10.1103/PhysRevA.86.053633} {\bibfield
  {journal} {\bibinfo  {journal} {Phys. Rev. A}\ }\textbf {\bibinfo {volume}
  {86}},\ \bibinfo {pages} {053633} (\bibinfo {year} {2012})}\BibitemShut
  {NoStop}%
\bibitem [{\citenamefont {von Keyserlingk}\ and\ \citenamefont
  {Conduit}(2013)}]{VonKeyserlingk2013}%
  \BibitemOpen
  \bibfield  {author} {\bibinfo {author} {\bibfnamefont {C.~W.}\ \bibnamefont
  {von Keyserlingk}}\ and\ \bibinfo {author} {\bibfnamefont {G.~J.}\
  \bibnamefont {Conduit}},\ }\href {\doibase 10.1103/PhysRevB.87.184424}
  {\bibfield  {journal} {\bibinfo  {journal} {Phys. Rev. B}\ }\textbf {\bibinfo
  {volume} {87}},\ \bibinfo {pages} {184424} (\bibinfo {year}
  {2013})}\BibitemShut {NoStop}%
\bibitem [{\citenamefont {Leggett}(2006)}]{Leggett2006}%
  \BibitemOpen
  \bibfield  {author} {\bibinfo {author} {\bibfnamefont {A.~J.}\ \bibnamefont
  {Leggett}},\ }\href@noop {} {\emph {\bibinfo {title} {{Quantum Liquids: Bose
  Condensation and Cooper Pairing in Condensed-Matter Systems}}}}\ (\bibinfo
  {publisher} {Oxford University Press},\ \bibinfo {address} {Oxford, UK},\
  \bibinfo {year} {2006})\BibitemShut {NoStop}%
\bibitem [{\citenamefont {Tan}(2008{\natexlab{a}})}]{Tan2008}%
  \BibitemOpen
  \bibfield  {author} {\bibinfo {author} {\bibfnamefont {S.}~\bibnamefont
  {Tan}},\ }\href {\doibase 10.1016/j.aop.2008.03.005} {\bibfield  {journal}
  {\bibinfo  {journal} {Ann. Phys. (NY)}\ }\textbf {\bibinfo {volume} {323}},\
  \bibinfo {pages} {2971} (\bibinfo {year} {2008}{\natexlab{a}})}\BibitemShut
  {NoStop}%
\bibitem [{\citenamefont {Tan}(2008{\natexlab{b}})}]{Tan2008a}%
  \BibitemOpen
  \bibfield  {author} {\bibinfo {author} {\bibfnamefont {S.}~\bibnamefont
  {Tan}},\ }\href {\doibase 10.1016/j.aop.2008.03.003} {\bibfield  {journal}
  {\bibinfo  {journal} {Ann. Phys. (NY)}\ }\textbf {\bibinfo {volume} {323}},\
  \bibinfo {pages} {2987} (\bibinfo {year} {2008}{\natexlab{b}})}\BibitemShut
  {NoStop}%
\bibitem [{\citenamefont {Tan}(2008{\natexlab{c}})}]{Tan2008b}%
  \BibitemOpen
  \bibfield  {author} {\bibinfo {author} {\bibfnamefont {S.}~\bibnamefont
  {Tan}},\ }\href {\doibase 10.1016/j.aop.2008.03.004} {\bibfield  {journal}
  {\bibinfo  {journal} {Ann. Phys. (NY)}\ }\textbf {\bibinfo {volume} {323}},\
  \bibinfo {pages} {2952} (\bibinfo {year} {2008}{\natexlab{c}})}\BibitemShut
  {NoStop}%
\bibitem [{\citenamefont {Braaten}\ \emph {et~al.}(2008)\citenamefont
  {Braaten}, \citenamefont {Kang},\ and\ \citenamefont
  {Platter}}]{Braaten2008}%
  \BibitemOpen
  \bibfield  {author} {\bibinfo {author} {\bibfnamefont {E.}~\bibnamefont
  {Braaten}}, \bibinfo {author} {\bibfnamefont {D.}~\bibnamefont {Kang}}, \
  and\ \bibinfo {author} {\bibfnamefont {L.}~\bibnamefont {Platter}},\ }\href
  {\doibase 10.1103/PhysRevA.78.053606} {\bibfield  {journal} {\bibinfo
  {journal} {Phys. Rev. A}\ }\textbf {\bibinfo {volume} {78}},\ \bibinfo
  {pages} {053606} (\bibinfo {year} {2008})}\BibitemShut {NoStop}%
\bibitem [{\citenamefont {Landau}\ and\ \citenamefont
  {Lifshitz}(2001)}]{Landau2001}%
  \BibitemOpen
  \bibfield  {author} {\bibinfo {author} {\bibfnamefont {L.}~\bibnamefont
  {Landau}}\ and\ \bibinfo {author} {\bibfnamefont {E.~M.}\ \bibnamefont
  {Lifshitz}},\ }\href@noop {} {\emph {\bibinfo {title} {{Statistical Physics.
  Part 1}}}}\ (\bibinfo  {publisher} {Butterworth-Heinemann},\ \bibinfo
  {address} {Oxford, UK},\ \bibinfo {year} {2001})\BibitemShut {NoStop}%
\bibitem [{\citenamefont {Friedman}\ and\ \citenamefont
  {Pandharipande}(1981)}]{Friedman1981}%
  \BibitemOpen
  \bibfield  {author} {\bibinfo {author} {\bibfnamefont {B.}~\bibnamefont
  {Friedman}}\ and\ \bibinfo {author} {\bibfnamefont {V.}~\bibnamefont
  {Pandharipande}},\ }\href {\doibase 10.1016/0375-9474(81)90649-7} {\bibfield
  {journal} {\bibinfo  {journal} {Nucl. Phys. A}\ }\textbf {\bibinfo {volume}
  {361}},\ \bibinfo {pages} {502} (\bibinfo {year} {1981})}\BibitemShut
  {NoStop}%
\bibitem [{\citenamefont {Lee}\ \emph {et~al.}(2011)\citenamefont {Lee},
  \citenamefont {Conduit}, \citenamefont {Nemec}, \citenamefont {{L\'{o}pez
  R\'{\i}os}},\ and\ \citenamefont {Drummond}}]{Lee2011}%
  \BibitemOpen
  \bibfield  {author} {\bibinfo {author} {\bibfnamefont {R.~M.}\ \bibnamefont
  {Lee}}, \bibinfo {author} {\bibfnamefont {G.~J.}\ \bibnamefont {Conduit}},
  \bibinfo {author} {\bibfnamefont {N.}~\bibnamefont {Nemec}}, \bibinfo
  {author} {\bibfnamefont {P.}~\bibnamefont {{L\'{o}pez R\'{\i}os}}}, \ and\
  \bibinfo {author} {\bibfnamefont {N.~D.}\ \bibnamefont {Drummond}},\ }\href
  {\doibase 10.1103/PhysRevE.83.066706} {\bibfield  {journal} {\bibinfo
  {journal} {Phys. Rev. E}\ }\textbf {\bibinfo {volume} {83}},\ \bibinfo
  {pages} {066706} (\bibinfo {year} {2011})}\BibitemShut {NoStop}%
\bibitem [{\citenamefont {Drummond}\ \emph {et~al.}(2008)\citenamefont
  {Drummond}, \citenamefont {Needs}, \citenamefont {Sorouri},\ and\
  \citenamefont {Foulkes}}]{Drummond2008}%
  \BibitemOpen
  \bibfield  {author} {\bibinfo {author} {\bibfnamefont {N.~D.}\ \bibnamefont
  {Drummond}}, \bibinfo {author} {\bibfnamefont {R.~J.}\ \bibnamefont {Needs}},
  \bibinfo {author} {\bibfnamefont {A.}~\bibnamefont {Sorouri}}, \ and\
  \bibinfo {author} {\bibfnamefont {W.~M.~C.}\ \bibnamefont {Foulkes}},\ }\href
  {\doibase 10.1103/PhysRevB.78.125106} {\bibfield  {journal} {\bibinfo
  {journal} {Phys. Rev. B}\ }\textbf {\bibinfo {volume} {78}},\ \bibinfo
  {pages} {125106} (\bibinfo {year} {2008})}\BibitemShut {NoStop}%
\bibitem [{\citenamefont {Yang}(1962)}]{Yang1962}%
  \BibitemOpen
  \bibfield  {author} {\bibinfo {author} {\bibfnamefont {C.~N.}\ \bibnamefont
  {Yang}},\ }\href {\doibase 10.1103/RevModPhys.34.694} {\bibfield  {journal}
  {\bibinfo  {journal} {Rev. Mod. Phys.}\ }\textbf {\bibinfo {volume} {34}},\
  \bibinfo {pages} {694} (\bibinfo {year} {1962})}\BibitemShut {NoStop}%
\bibitem [{\citenamefont {Annett}(2004)}]{Annett2004}%
  \BibitemOpen
  \bibfield  {author} {\bibinfo {author} {\bibfnamefont {J.~F.}\ \bibnamefont
  {Annett}},\ }\href@noop {} {\emph {\bibinfo {title} {{Superconductivity,
  Superfluids and Condensates}}}}\ (\bibinfo  {publisher} {Oxford University
  Press},\ \bibinfo {address} {Oxford, UK},\ \bibinfo {year}
  {2004})\BibitemShut {NoStop}%
\bibitem [{\citenamefont {Eagles}(1969)}]{Eagles1969}%
  \BibitemOpen
  \bibfield  {author} {\bibinfo {author} {\bibfnamefont {D.~M.}\ \bibnamefont
  {Eagles}},\ }\href {\doibase 10.1103/PhysRev.186.456} {\bibfield  {journal}
  {\bibinfo  {journal} {Phys. Rev.}\ }\textbf {\bibinfo {volume} {186}},\
  \bibinfo {pages} {456} (\bibinfo {year} {1969})}\BibitemShut {NoStop}%
\bibitem [{\citenamefont {{De Palo}}\ \emph {et~al.}(2002)\citenamefont {{De
  Palo}}, \citenamefont {Rapisarda},\ and\ \citenamefont
  {Senatore}}]{DePalo2002}%
  \BibitemOpen
  \bibfield  {author} {\bibinfo {author} {\bibfnamefont {S.}~\bibnamefont {{De
  Palo}}}, \bibinfo {author} {\bibfnamefont {F.}~\bibnamefont {Rapisarda}}, \
  and\ \bibinfo {author} {\bibfnamefont {G.}~\bibnamefont {Senatore}},\ }\href
  {\doibase 10.1103/PhysRevLett.88.206401} {\bibfield  {journal} {\bibinfo
  {journal} {Phys. Rev. Lett.}\ }\textbf {\bibinfo {volume} {88}},\ \bibinfo
  {pages} {206401} (\bibinfo {year} {2002})}\BibitemShut {NoStop}%
\bibitem [{\citenamefont {Ortiz}\ and\ \citenamefont
  {Dukelsky}(2005)}]{Ortiz2005}%
  \BibitemOpen
  \bibfield  {author} {\bibinfo {author} {\bibfnamefont {G.}~\bibnamefont
  {Ortiz}}\ and\ \bibinfo {author} {\bibfnamefont {J.}~\bibnamefont
  {Dukelsky}},\ }\href {\doibase 10.1103/PhysRevA.72.043611} {\bibfield
  {journal} {\bibinfo  {journal} {Phys. Rev. A}\ }\textbf {\bibinfo {volume}
  {72}},\ \bibinfo {pages} {043611} (\bibinfo {year} {2005})}\BibitemShut
  {NoStop}%
\end{thebibliography}%

\end{document}